\documentclass[preprint,aps,pre]{revtex4-1}
\usepackage{amsmath,amsfonts,amsthm,float,rotating,verbatim,subfigure}

\begin{document}
\title{Binding Energies of Charged Metal Nanoparticle Configurations}
\author{Alexander Moore}
\email{amm456@cornell.edu}
\affiliation{Department of Physics, University of Chicago}
\altaffiliation [Currently in]{ Field of Theoretical and Applied Mechanics, Cornell University}
\date{\today}
\begin{abstract}
The electrostatic interaction between metal spheres is an influential component in the assembly of many nanoscale materials in chemistry. Here we derive a method to calculate the energy and polarizations of metal spheres in arbitrary configurations to an arbitrary multipole order. This helps provide insight into the preferred configurations of charged metal particles and demonstrates the sensitivity of electrostatic interactions to configuration geometry. \end{abstract}
\maketitle
\section{Introduction}
One of the extensions of electrostatics beyond coulomb interactions is the interaction of charged conductors with multiple modes of polarization. The relatively simple problem of two charged metal spheres has been studied from a theoretical and pedagogical standpoint using image charges \cite{Soules} \cite{Slisko}. In addition, more complex methods using image charges  \cite{Phillips} and boundary element methods \cite{Kwon1998} have been developed for multiple-particle configurations.

Here we derive a new method to calculate the electrostatic interaction energy of charged conducting spheres that utilizes multipole expansions of the potential and charge density. This method can be expanded to any number of conductors or to an arbitrary precision by including higher order multipoles in the calculation. This approach is motivated by the assembly configurations of binary nanoparticle super lattices (BNSLs). Guided by a complex combination of entropic and electromagnetic interactions, these lattices can assemble in a vast array of different shapes and configurations. In order to better understand the assembly of these lattices, a number of experimental studies have tried to isolate and analyze the different interactions driving the assembly\cite{Shevchenko}. Here we focus on analyzing the electrostatic interactions that guide the assembly of the metal nanoparticles at the center of these lattices.

Normally in electrostatic problems, most of the higher order interactions can be ignored since the monopole and dipole terms dominate the energy equation for large separations. Once the scale of the configuration approaches the scale of the nanoparticles, the quadrupole or octupole interactions can be on the same order as some monopole interactions \cite{Talapin}. Since these assemblies can also include nanoparticles in contact, there is no assumption that can justify ignoring multipole terms beyond the dipole. This multipole expansion method allows the interaction between any number of conducting spheres to be calculated using an arbitrary number of multipoles. This will provide insight into why BNSLs only assemble with certain configurations of metal nanoparticles by allowing a direct energy comparison across different configurations. Moreover, it will allow us to analyze how the geometry of a configuration influences the relative strength of various multipoles in a configuration. \\
\indent We proceed in deriving the equations for the energy of the configuration by first looking at the energy between two charged spheres. To make this calculation easier, we begin by assuming that the two spheres lie on the z-axis of our arbitrarily selected coordinate system. Once we find the energy equations in this simplified aligned case, we derive the transformation to express the energy equations in any arbitrary frame. The accuracy of this method is then checked against existing literature calculations, and then we generalize the energy equations to include an arbitrary number of spheres. Finally, we present calculations for a number configurations and comment on the accuracy and insights of the multipole expansion method.
\section{Method and Derivation}
To calculate the multipole expansion of the energy, we start with the energy equation:
\begin{equation} E=\frac{1}{2}\int \limits_{All Space} \rho \thinspace V\mathrm{d^3}x \end{equation}
This can be broken up into the pairwise sphere interactions for N spheres:
\begin{equation} \label{eq:ijenergy}
E=\frac{1}{2} \sum \limits_{i=1}^{N} \sum \limits_{j=1}^{N} \int \limits_{\text{Sph i}} \rho_{i} V_{j}\mathrm{d^3}x
\end{equation}
where $\rho_i$ is the charge density on sphere i, $V_j$ is the potential on sphere j, and the integration is carried over the volume of sphere i. The terms where $i=j$ represent the energy cost of holding a charge distribution on the sphere, further referred to as the self-energy. The other terms represent the interaction energy, or the energy associated with charge distributions and potentials on different spheres. We now find the multipole expansion of the charge distribution and potential for a general sphere $S$. The charge distribution on sphere $S$ (with radius $r_S$) is given by:
\begin{equation} \label{eq:dist}
\rho_{S}=\sum_{\ell,m} S_{\ell}^{m*} Y_{\ell}^{m*} (\theta,\phi) \delta (r-r_{S}) 
\end{equation}
where $S_{\ell}^{m*}, Y_{\ell}^{m*} (\theta,\phi)$, etc. represent the complex conjugates of $S_{\ell}^{m}, Y_{\ell}^{m} (\theta,\phi),$ etc. The $S_{\ell}^{m}$'s are related to the multipole moments, $q_{\ell m}$'s, as defined by Jackson (Gaussian or cgs units) \cite[p.145]{Jackson} where:
\begin{equation} 
S_{\ell}^{m}=\frac{q_{\ell m}}{r_{S}^{\ell+2}}=\frac{1}{r_{S}^{\ell+2}}\int Y_{\ell}^{m*}(\theta,\phi)r^{\ell} \rho_S(\vec{x}) \mathrm{d^3}x
\end{equation}
To calculate the potential, we expand the formula for the potential in terms of the sphere's multipoles:
\begin{equation} \label{eq:potent}
V_S(\vec{r}) = \int \frac{\rho_S(\vec{r}')}{|\vec{r} - \vec{r}'|}\mathrm{d^3}r'=\sum_{\ell,m} \frac{4\pi}{2\ell+1} r_{S}^{\ell+2} S_{\ell}^{m} \frac{Y_{\ell}^{m}(\theta,\phi)}{|\vec{r}|^{\ell+1}}
\end{equation}
\subsection{Z-Axis Aligned Case}
To calculate the energy explicitly, we start by just considering two spheres, A and B, with their charge distributions and potentials given by replacing the label S with labels A and B in equations \ref{eq:dist} and \ref{eq:potent} . Thus the total interaction energy is given by:
\begin{figure}
\centering
\includegraphics [height=80mm] {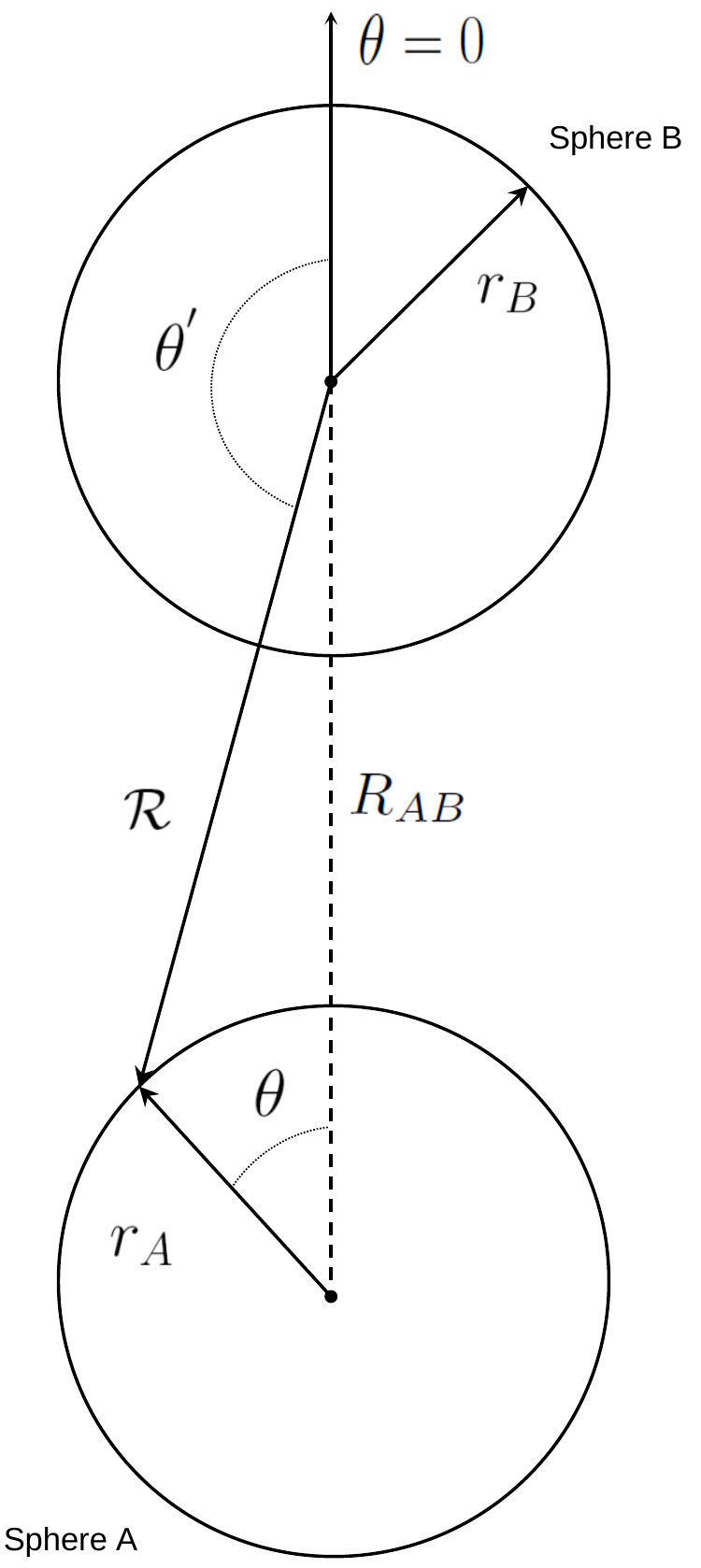}
\caption{Diagram of the two spheres in the aligned case.}
\label{fig:diag1}
\end{figure}

\begin{align}
E^{int} & = \frac{1}{2} \int \limits_{A}\rho_{A} V_{B}\mathrm{d^3}x+ \frac{1}{2}\int \limits_{B}\rho_{B} V_{A}\mathrm{d^3}x \\
& = E^{AB} + E ^{BA}
\end{align}

\indent We will continue by deriving the formula for $E^{AB}$. The other half of the interaction energy, $E^{BA}$, can be derived by using $\rho_B$ and $V_A$ in place of $\rho_A$ and $V_B$\footnote{When deriving the form for $E^{BA}$, all odd order multipoles need to be multiplied by a $-1$ since the alignment of a positive multipole  This adds a $(-1)^{\ell+\ell^{'}}$ factor to the odd ordered B multipoles in the derivation of $E^{BA}$. }. For the calculation of two aligned spheres, the radius term in the potential ($\mathcal{R} $ in eq. \ref{eq:potent}) is the distance between a point on sphere A and the center of sphere B. This gives the $E^{AB}$ interaction energy as:
\begin{equation} E^{AB} = \sum_{\ell,\ell^{'},m,m^{'}} \frac{1}{2}\int \frac{4\pi}{2\ell^{'}+1} r_{B}^{\ell^{'}+2} A_{\ell}^{m*} B_{\ell^{'}}^{m^{'}} \frac{Y_{\ell}^{m*} (\theta,\phi)Y_{\ell^{'}}^{m^{'}}(\theta^{'},\phi^{'} )}{\mathcal{R}^{\ell^{'}+1}}\delta (r-r_{A})\mathrm{d^3}x \end{equation}
We can find the form for $\mathcal{R}$ using the law of cosines:
\begin{equation}
\mathcal{R}=\sqrt{(R_{AB})^{2}+(r_{A})^2-2(R_{AB})(r_{A})\cos{\theta}}^{\thinspace \ell^{'}+1}
\end{equation}
Plugging this in and integrating over the radius variable, gives the energy as: 
\begin{equation} \label{eq:integral}
E^{AB}=\sum_{\ell,\ell^{'},m,m^{'}} \frac{2\pi}{2\ell^{'}+1} A_{\ell}^{m*} B_{\ell^{'}}^{m^{'}} \frac{r_{A}^{2}\thinspace r_{B}^{\ell^{'}+2}}{R_{AB}^{\ell^{'}+1}}\int \frac{Y_{\ell}^{m*} (\theta,\phi)Y_{\ell^{'}}^{m^{'}}(\theta^{'},\phi^{'} )}{\sqrt{1+(\frac{r_{A}}{R_{AB}})^2-2(\frac{r_{A}}{R_{AB}})\cos{\theta}}^{\thinspace \ell^{'}+1}}\mathrm{d}\Omega 
\end{equation}
The expression contained inside the radical is the generating function for the Gegenbauer Polynomials\cite{Abramowitz}
\begin{equation} \sqrt{1-2xz+z^{2}}^{\thinspace -2 \zeta}=\sum \limits_{n=0}^{\infty} C_{n}^{(\zeta)}(x)z^{n} \quad , \quad |z|<1 \end{equation}
with $z= \frac{r_A}{R_{AB}}$ , $ x=\cos{\theta}$, and $\zeta=\frac{\ell^{'}+1}{2}$. This allows us to express the denominator as a power series in $(r_{A}/R_{AB})$ with the coefficients being the Gegenbauer polynomials$(C_{n}^{(\zeta)}(x))$.
Through a transformation matrix of coefficients, the denominator is changed from an expansion of Gegenbauer polynomials into an expansion of spherical harmonics. This is done through the following transformation:
\begin{align} C_{n}^{(\zeta)}(\cos{\theta}) & = \sum \limits_{k=0}^{n} c_{nk}^{\zeta} \cos^k(\theta) = \sum \limits_{k,p=0}^{n} c_{nk}^{\zeta} h_{kp} P_{p}(\cos{\theta}) \\
& = \sum \limits_{k,p=0}^{n} c_{nk}^{\zeta} h_{kp} \sqrt{\frac{4\pi }{2p+1}} Y_{p}^{0}(\theta, \phi) \\
& = \sum \limits_{p =0}^{n}\mathcal {T}_{np}^{\zeta}Y_{p}^{0}(\theta, \phi) \end{align}
where we defined $\mathcal {T}_{np}^{\zeta}$ as:
\begin{equation}
\mathcal {T}_{np}^{\zeta}=\sum \limits_{k=0}^{n} c_{nk}^{\zeta} h_{kp} \sqrt{\frac{4\pi }{2p+1}}
\end{equation}
and utilized the following identity:
\begin{equation}
\sum \limits_{p=0}^{n} h_{kp} P_{p}(\cos{\theta})=(cos{\theta})^{k}
\end{equation}
where $P_{p}(\cos{\theta})$ is the p-th Legendre polynomial \cite[p. 798]{Abramowitz}. \\
\indent The formula for $\theta^{'}$ is derived from the law of sines: $\frac{\sin{(\pi-\theta^{'})}}{r_{A}}=\frac{\sin{\theta}}{\mathcal{R}}$ (see figure \ref{fig:diag1}). As before, we can expand $\mathcal{R}$ using the law of cosines and the Gegenbauer series expansion:
\begin{align}
\theta^{'}&=\pi-\arcsin{\left(\frac{r_{A}\sin{\theta}}{R_{AB}\sqrt{1+(\frac{r_{A}}{R_{AB}})^2-2(\frac{r_{A}}{R_{AB}})\cos{\theta}}}\right)} \\
\theta^{'} & =\pi-\arcsin{\left(\frac{r_{A}\sin{\theta}}{R_{AB}}\sum \limits_{t}[C_{t}^{(\frac{1}{2})}(\cos{\theta})(\frac{r_{A}}{R_{AB}})^{t}]\right)}
\end{align}
where $C_{t}^{(\frac{1}{2})}$ is the t-th Gegenbauer polynomial. Because of the azimuthal symmetry of the set-up we see that $\phi '=\phi$. Plugging this all into the energy equation (leaving $\theta '$ in its place for simplicity):
\begin{equation} E^{AB}=\sum_{\ell,\ell^{'},m,m^{'}} \mathcal{K}_{\ell,\ell^{'}}^{m,m^{'}} \int Y_{\ell}^{m*} (\theta,\phi)Y_{\ell^{'}}^{m^{'}}(\theta^{'},\phi) \sum \limits_{n=0}^{\infty}(\frac{r_{A}}{R_{AB}})^{n}\sum \limits_{p=0}^{n} [\mathcal{T}_{np}^{\zeta} Y_{p}^{0} (\theta,\phi)]\mathrm{d}\Omega \end{equation}
\begin{equation} E^{AB}= \sum_{\ell,\ell^{'},m,m^{'}} \mathcal{K}_{\ell,\ell^{'}}^{m,m^{'}} \sum \limits_{n=0}^{\infty}(\frac{r_{A}}{R_{AB}})^{n} \sum \limits_{p=0}^{n} \mathcal{T}_{np}^{\zeta}\int {Y_{\ell}^{m*} (\theta,\phi)Y_{p}^{0} (\theta,\phi)Y_{\ell^{'}}^{m^{'}}(\theta^{'},\phi)} \mathrm{d}\Omega \end{equation}
where $\mathcal{K}_{\ell,\ell^{'}}^{m,m^{'}}$ is defined as:
\begin{equation} \mathcal{K}_{\ell,\ell^{'}}^{m,m^{'}}= \frac{2\pi}{2\ell^{'}+1} \frac{r_{A}^{2}\thinspace r_{B}^{\ell^{'}+2}}{R_{AB}^{\ell^{'}+1}} A_{\ell}^{m*} B_{\ell^{'}}^{m^{'}}\end{equation}
and $\zeta=\frac{\ell^{'}+1}{2}$. We can also simplify the integral to be performed by utilizing the Wigner 3-j symbols to decompose the product of spherical harmonics into a sum. This formula is given by \cite[p.1057]{Messiah}:
\begin{align} Y_{\ell}^{m*} (\Omega)Y_{p}^{0} (\Omega) &= \sum \limits_{L=|\ell-p|}^{\ell+p} \sqrt{\frac{(2 \ell + 1) (2 p + 1) (2 L + 1)}{4\pi}} \begin{pmatrix} \ell & p & L \\ m & 0 & -m \end{pmatrix} Y_{L}^{m*}(\Omega) \begin{pmatrix} \ell & p & L \\ 0 & 0 & 0 \end{pmatrix}\\
& = \sum \limits_{L=|\ell-p|}^{\ell+p} W^{L m}_{\ell, p} \thinspace \thinspace Y_{L}^{m*}(\Omega)
\end{align}
where $\begin{pmatrix} \ell & p & L \\ \ell_{z} & p_{z} & L_{z} \end{pmatrix}$ is the appropriate Wigner 3-j symbol with the convention used in \cite[p.1053]{Messiah} and $\Omega=(\theta,\phi)$. This gives us a final expression for $E^{AB}$:
\begin{equation} \label{eq:f_aligned}
E^{AB}=\sum \limits_{\ell,\ell^{'},m,m^{'}} \mathcal{K}_{\ell,\ell^{'}}^{m,m^{'}} \sum \limits_{n=0}^{\infty}(\frac{r_{A}}{R_{AB}})^{n} \sum \limits_{p=0}^{n} \mathcal{T}_{np}^{\zeta} \sum \limits_{L=|\ell-p|}^{\ell+p} \thinspace W^{L m}_{\ell,p} \int {Y_{L}^{m*}(\theta,\phi)Y_{\ell^{'}}^{m^{'}}(\theta^{'},\phi)} \mathrm{d}\Omega \end{equation}
 This is the interaction energy that we will use to calculate the energy for a configuration of spheres. To give an example of what the individual energy terms are, we integrate this formula explicitly for the first few multipoles. For simplicity in reporting the expansion below, we assume that the spheres are of equal size ($r_A = r_B = r$). In addition, we will assume the two spheres are separated by a distance much larger than their individual radii ($\frac{r}{R_{AB}}<<1$). This allows us to take the following simplification \footnote{This makes the interactions much simpler by eliminating many of the azimuthal $m \neq 0$ multipoles that occur in the general case interaction energy. }:
\begin{equation} \theta^{'}=\pi-\arcsin{\left(\frac{r_{A}\sin{\theta}}{R_{AB}}\sum \limits_{t}[C_{t}^{(\frac{1}{2})}(\cos{\theta})(\frac{r_{A}}{R_{AB}})^{t}]\right)} \rightarrow \pi \end{equation}
Using this simplification we expand $E^{AB}$ and report the terms of order $r^3(\frac{r}{R_{AB}})^3$ or lower:
\begin{align}E^{AB} = & \frac{2 \thinspace \pi \thinspace r^{4}\thinspace A_{0}^{0} B_{0}^{0}}{R_{AB}}+\frac{2\thinspace \pi \thinspace r^{5}\thinspace A_{1}^{0} B_{0}^{0}}{\sqrt{3}(R_{AB})^{2}}-\frac{2\thinspace \pi \thinspace r^{5}\thinspace A_{0}^{0} B_{1}^{0}}{\sqrt{3}(R_{AB})^{2}}-\frac{4\thinspace \pi \thinspace r^{6}\thinspace A_{1}^{0} B_{1}^{0}}{3(R_{AB})^{3}} \nonumber \\ & \quad \quad+ \frac{2\thinspace \pi \thinspace r^{4}\thinspace A_{2}^{0} B_{0}^{0}}{\sqrt{5}(R_{AB})^{3}} + \frac{2\thinspace \pi \thinspace r^{6}\thinspace A_{0}^{0} B_{2}^{0}}{\sqrt{5}(R_{AB})^{3}}\end{align}
These are the terms of the multipole expansion of the partial interaction energy, $E^{AB}$, up to the dipole-dipole interaction in the special case where the spheres are far apart and aligned along the $\theta=0$ axis.


\begin{figure}
\centering
\subfigure[Three dimensional perspective]{
\includegraphics[width=2.9in] {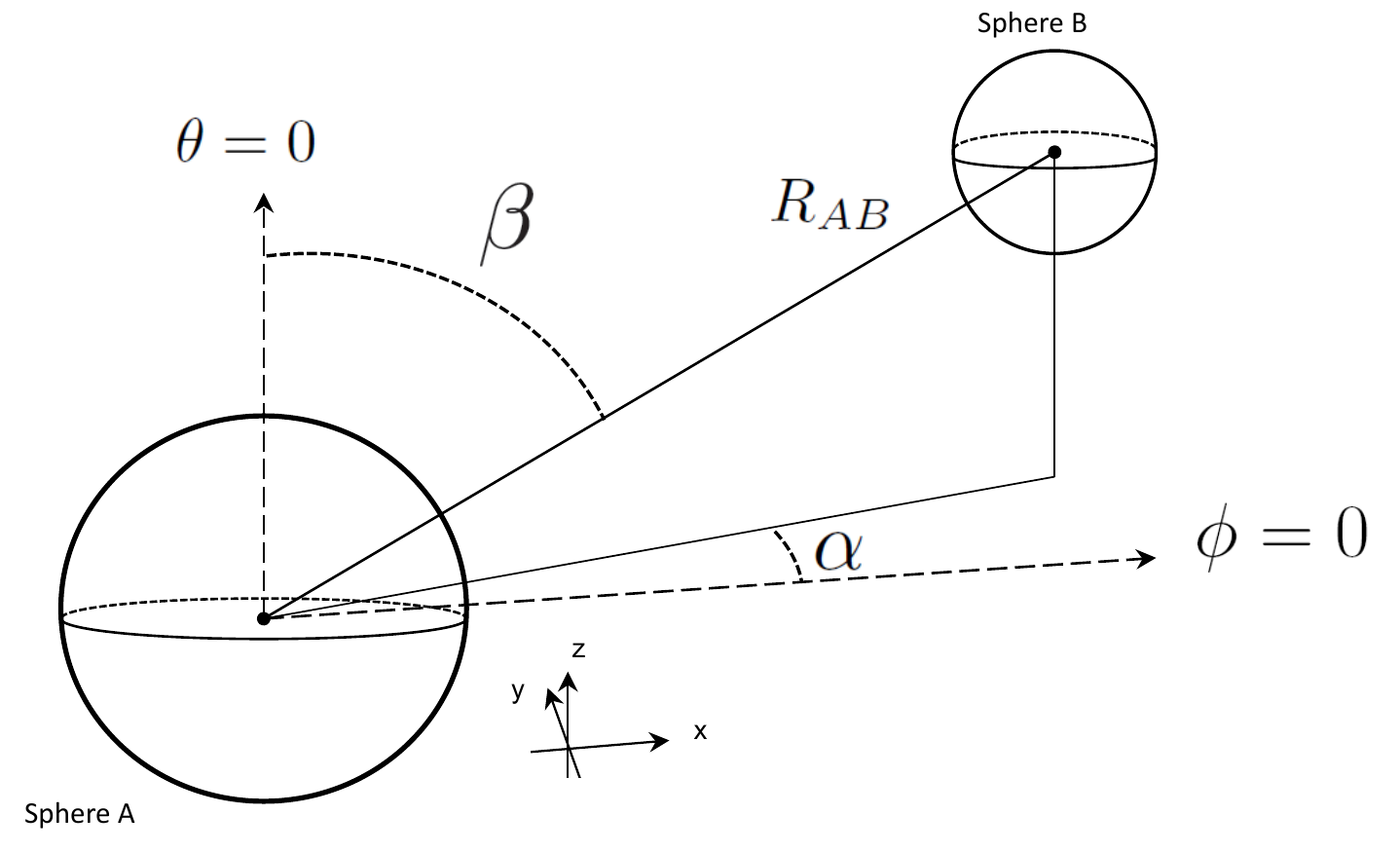}
\label{fig:subfig2.1}}
\subfigure[x-z plane]{
\includegraphics[width=2.5in] {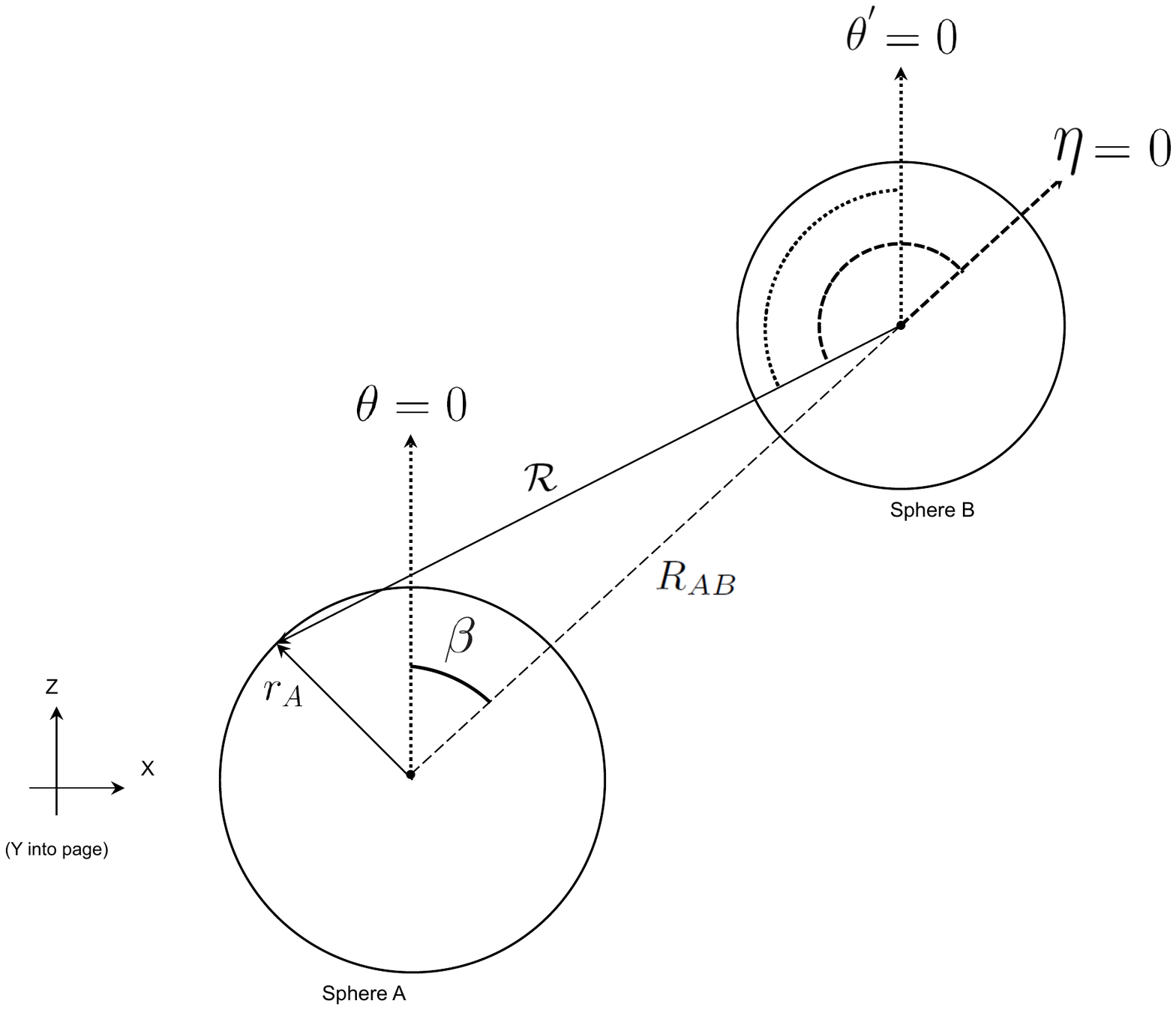}
\label{fig:subfig2.2}}
\caption{Non-aligned configuration with Euler angles $\alpha$ and $\beta$ marked. \quad \quad}
\label{fig:2}
\end{figure}

\subsection{Non-Aligned Case}
\label{sec:nonaligned}

To generalize for spheres not aligned along the $\theta = 0$ axis, the expression for $\mathcal{R}$ becomes:
\begin{align}\mathcal{R}^{-(\ell^{'}+1)} & =\sqrt{(R_{AB})^{2}+(r_{A})^2-2(R_{AB})(r_{A})\cos{\eta}}^{\thinspace-( \ell^{'}+1)} \nonumber \\
& = R_{AB}^{-(\ell^{'}+1)}\sum \limits_{n=0}^{\infty} (\frac{r_{A}}{R_{AB}})^{n} \sum \limits_{p=0}^{n} [\mathcal{T}_{np}^{\zeta} Y_{p}^{0} (\eta,\tau)]\end{align}
where $\eta$ and $\tau$ are the polar and azimuthal angles between $R_{AB}$ and the vector $\mathcal{R}$. When the spheres were aligned along the $z=0$ axis, we were able to simplify the equations since $\eta=\theta$ and $\tau=\phi$.\\
\indent We start with same Gegenbauer expansion method presented above, but we also need to rotate the spherical harmonics in $(\eta, \tau)$ to spherical harmonics in $(\theta,\phi)$ by using the Euler angles $(\alpha, \beta, \gamma)$ that separate the two frames. In our specific case, we are only rotating the center point of a sphere to the z-axis without changing the orientation. This means that one of the Euler angles, $\gamma$, will not have any effect on the orientation of the spheres. Thus we can take $\gamma=0$ and just use two Euler angles to describe the rotation. The following transformation converts the spherical harmonics with $(\eta, \tau)$ arguments to spherical harmonics with $(\theta,\phi)$ arguments:
\begin{equation} Y_{\ell}^{m} (\eta,\tau) = \sum_{M=-\ell}^{\ell} \mathcal{D}_{M,m}^{\ell}(0,-\beta,-\alpha,)\thinspace Y_{\ell}^{M} (\theta,\phi) \end{equation}
where $\mathcal{D}_{M,m}^{\ell}$ is the standard finite rotations operator that can be found in Landau and Lifshitz \cite[p.218]{Landau}.\\
\indent This transformation allows us to easily calculate the interaction energy equation for a configuration of multiple spheres. In a configuration with more than two spheres, each pairwise interaction energy has the same form as equation \ref{eq:f_aligned} when the coordinate systems are aligned along each line of action (i.e. along $R_{AB}$ as $\eta$ is in Figure 2b.). Instead of recalculating the entire energy series for every orientation, we can utilize the fact that the rotation of the spherical harmonics is a linear transformation (see Appendix \ref{app:linear}). The transformation allows us to take each multipole referred in these local pairwise frames and transform them all to the same configuration-wide $(\theta,\phi)$ coordinate system. This expands the $A^{m}_{\ell}$ and $B^{m'}_{\ell^{'}}$ multipoles in $(\eta,\tau)$ in terms of new multipoles $\hat{A}^{m}_{\ell}$ and $\hat{B}^{m'}_{\ell^{'}}$ in the $(\theta,\phi)$ frame. This gives us the transformation:
\begin{equation} \Longrightarrow A^{M}_{\ell}=\sum \limits_{M=-\ell}^{\ell} \mathcal{D}_{M,m}^{\ell}(0,-\beta,-\alpha,)\hat{A}^{M}_{\ell} \end{equation}
The rotation coefficients can be further decomposed into their three constituent parts:
\begin{equation}\mathcal{D}_{M,m}^{\ell}(0,-\beta,-\alpha,)=e^{iM \cdot 0}d^{\ell}_{M,m}(-\beta)e^{-iM\alpha}
\end{equation}
\begin{eqnarray} A^{m}_{\ell} & =\sum \limits_{M=-\ell}^{\ell} d^{\ell}_{M,m}(-\beta)\hat{A}^{M}_{\ell} e^{-iM\alpha} \\
A^{m *}_{\ell} & =\sum \limits_{M=-\ell}^{\ell} d^{\ell}_{M,m}(-\beta)\hat{A}^{M*}_{\ell} e^{iM\alpha}\end{eqnarray}
Applying this transformation to the multipoles changes the expression for $\mathcal{K}$ to:
\begin{equation} \mathcal{K}_{\ell,\ell^{'}}^{m,m^{'}}= \frac{2\pi}{2\ell^{'}+1} \frac{r_{A}^{2}\thinspace r_{B}^{\ell^{'}+2}}{R_{AB}^{\ell^{'}+1}} \left( \sum \limits_{M=-\ell}^{\ell} d^{\ell}_{M,m}(-\beta)\hat{A}^{M*}_{\ell} e^{iM\alpha}\right)\left( \sum \limits_{M'=-\ell^{'}}^{\ell^{'}} d^{\ell^{'}}_{M',m'}(-\beta)\hat{B}^{M'}_{\ell^{'}} e^{-iM'\alpha}\right)\end{equation}
This new form for $\mathcal{K}$ is then plugged into equation \ref{eq:f_aligned}. This gives us a completely general expression for the interaction energy between two arbitrarily oriented spheres. To see what the first few energy terms are, we integrate equation \ref{eq:f_aligned} explicitly and obtain the formula for the interaction energy between the two spheres. Due to limited space we show only the monopole-monopole and dipole-monopole terms in the $\frac{r}{R_AB}<<1$ limit below:
\begin{eqnarray}
E^{AB} &=& \frac{2 \pi \thinspace r_{A}^{2}\thinspace r_{B}^{2}A_{0}^{0} B_{0}^{0}}{R_{AB}}+\frac{2\sqrt{\frac{2}{3}}\pi\cos{(\frac{\beta}{2})}\sin{(\frac{\beta}{2})}r_{A}^{3}\thinspace r_{B}^{2}A_{1}^{-1*} B_{0}^{0}e^{-i\alpha}}{(R_{AB})^{2}} \\ \nonumber
& &\thinspace + \thinspace \frac{2\thinspace \pi (\cos{(\frac{\beta}{2})}^{2} - \sin{(\frac{\beta}{2})}^{2})\thinspace r_{A}^{3}\thinspace r_{B}^{2}A_{1}^{0} B_{0}^{0}}{\sqrt{3}(R_{AB})^{2}} \thinspace +\thinspace \frac{2\sqrt{\frac{2}{3}}\pi\cos{(\frac{\beta}{2})}\sin{(\frac{\beta}{2})}r_{A}^{3}\thinspace r_{B}^{2}A_{1}^{1*} B_{0}^{0}e^{i\alpha}}{(R_{AB})^{2}}+...
\end{eqnarray}


\subsection{Self-Energy}

The self-energy formulas are much easier to derive. Starting with the expression for the energy in equation \ref{eq:ijenergy}, we set $i=j$ and use the charge density and potential definitions from equations \ref{eq:dist} and \ref{eq:potent}. This gives the self-energy for sphere A is:
\begin{equation} E^{self} = \frac{1}{2} \sum \limits_{\ell,\ell^{'},m,m^{'}} \int \frac{4\pi}{2\ell^{'}+1} r_{A}^{\ell'+2} A_{\ell}^{m*} A_{\ell^{'}}^{m^{'}} \frac{Y_{\ell}^{m*} (\theta,\phi)Y_{\ell^{'}}^{m^{'}}(\theta,\phi)}{r^{\ell^{'}+1}}\delta (r-r_{A})\mathrm{d^3}x \end{equation}
Simplifying and utilizing the orthogonality of the spherical harmonics to perform the integration yields:
\begin{equation} E^{self}=\sum_{\ell,m}\frac{2\thinspace \pi}{2\ell+1}r_{A}^{3}A_{\ell}^{m*}A_{\ell}^{m} \end{equation}

Applying the same process to sphere B yields the same exact result in terms of the $B_{\ell}^{m}$'s instead of the $A_{\ell}^{m}$'s. By combining the interaction energies and the self energy from both spheres we obtain a energy formula that can be expanded to any order interaction.


\subsection{Green's Theorem Simplification} \label{sec:green}
\indent Due to the many summations in the interaction energy formula, the computation of relatively small order interactions (like the monopole-dipole interaction) can be extremely complicated and asymmetric. For example, the computation time needed to calculate $E^{AB}_{m-d}$  (a monopole on sphere A and a dipole on sphere B)\footnote{Convention: $E^{AB}_{d-m}$ is the interaction energy between a dipole on sphere A and a monopole on sphere B calculated with the $E^{AB}$ formula. $E^{BA}_{m-q}$ is the interaction energy between a monopole on sphere B and a quadrupole on sphere A calculated from the $E^{BA}$ formula} is much longer than the computation time needed to calculate $E^{BA}_{d-m}$ even though both represent the same electrostatic interaction. Looking at eq. \ref{eq:integral}, it is clear that the $\ell'$ exponent makes $E^{AB}$ more difficult to calculate as the multipole order increases.

This calculation can be simplified by making use of Green's reciprocation theorem to reduce the order of the exponent in eq. \ref{eq:integral} \cite[p.52]{Jackson}. When constructing the interaction energy $E^{AB}$, we can split the calculation up into each of the different components:
\begin{equation}
E^{AB}= E^{AB}_{m-m}+E^{AB}_{d-m}+E^{AB}_{m-d}+E^{AB}_{d-d}+ ...
\end{equation}
From Green's reciprocation theorem, we know that:
\begin{equation}
E^{AB}_{m-d} = E^{BA}_{d-m}
\end{equation}
In order to calculate $E^{AB}_{m-d}$, we would need to expand $Y_{1}^{m'}(\theta^{'},\phi)$ in another series. Thus, it is advantageous to use $E^{BA}_{d-m}$ instead since it avoids having to use any further expansions or unnecessary computation time. This renders the interaction energy as:
\begin{equation}
E^{AB}= E^{AB}_{m-m}+E^{AB}_{d-m}+E^{BA}_{d-m}+E^{AB}_{d-d}+ ...
\end{equation}
This same idea is then applied to further terms in the expansion (e.g. replacing $E^{AB}_{d-q}$ with $E^{BA}_{q-d}$).\\
\indent In addition, Green's reciprocal theorem also states that $E^{AB} = E^{BA}$. With proper attention to the conjugation of multipoles, there is no need to calculate $E^{BA}$. Since substitutions can rotate our aligned energy into any rotated frame (see section \ref{sec:nonaligned}), no calculations beyond $E^{AB}$ in the aligned case are needed to obtain the full interaction energy ($E^{int}$), for any orientation.
\subsection{Checking the formula} \label{sec:check}
Here we will check specific terms of the energy equation against other known equations. We will use the simpler aligned version of the energy equation for these calculations. The first and easiest check is that the monopole-monopole interaction gives the correct coulomb energy. For two charged spheres, we use the definitions of the multipole moments in Jackson \cite[p.146]{Jackson}, $A_{0}^{0}=\frac{q_A}{r_{A}^{2}\sqrt{4\thinspace\pi}}$ to find:
\begin{align} E^{int}_{m}= & E^{AB}_{m-m}+E^{BA}_{m-m} = 2E^{AB}_{m-m} = \frac{4 \thinspace \pi \thinspace r_{A}^{2}\thinspace r_{B}^{2}A_{0}^{0} B_{0}^{0}}{R_{AB}}=\frac{4 \thinspace \pi \thinspace r_{A}^{2}\thinspace r_{B}^{2}(\frac{q_{A}}{r_{A}^{2}\sqrt{4\thinspace\pi}}) (\frac{q_{B}}{r_{B}^{2}\sqrt{4\thinspace\pi}})}{R_{AB}}=\frac{q_{A}\thinspace q_{B}}{R_{AB}} \end{align}
which agrees with the coulomb energy of two point charges.
The dipole-monopole and the dipole-dipole energies derived from our equations also agree with the Jackson formulas. To check the dipole-monopole energy, we use the simplification of section \ref{sec:green} and plug in the values of the multipoles to find: 
\begin{align}
E^{int}_{d-m}  &=E^{AB}_{d-m}+E^{BA}_{m-d}=2E^{AB}_{d-m} \\ 
& =\frac{4\thinspace \pi \thinspace r_{A}^{3}\thinspace r_{B}^{2}A_{1}^{0} B_{0}^{0}}{\sqrt{3}(R_{AB})^{2}}=\frac{4\thinspace \pi \thinspace r_{A}^{2}\thinspace r_{B}^{3}(\frac{p_{A}}{r_{A}^{3}\sqrt{\frac{4\thinspace\pi}{3}}})( \frac{q_{B}}{r_{B}^{2}\sqrt{4\thinspace\pi}})}{\sqrt{3}(R_{AB})^{2}}=p_{A}\frac{q_{B}}{(R_{AB})^{2}}=p_{A}\vec{E}_{field}
\end{align}
where $p_{A}$ is the dipole on sphere A, and $\vec{E}_{field}$ is the electric field due to a monopole on sphere B. This agrees with the formula for the energy of a perfect dipole in an electric field.
For the dipole-dipole interaction, we start with the formula for the interaction between two dipoles given by Jackson \cite[p.147]{Jackson}:
\begin{equation} E_{dip,dip}=\frac{\vec{p}_{1}\cdotp \vec{p}_{2}-3(\vec{n} \cdotp \vec{p}_{1})(\vec{n} \cdotp \vec{p}_{2})}{|\vec{x}_{1}-\vec{x}_{2}|^{3}} \end{equation}
Since this formula is for the energy between two perfect dipoles, we recalculate the dipole-dipole term of our interaction energy assuming $\frac{r_A}{R_{AB}}<<1$ in order to compare it to the Jackson formula. Using the Green's theorem simplification from section \ref{sec:green} again and taking the dipole-dipole term from the $E^{AB}$ expansion yields:
\begin{align}
E^{int}_{d-d} & = E^{AB}_{d-d} + E^{BA}_{d-d} = 2E^{AB}_{d-d} 
 = \frac{8\thinspace \pi r_{A}^{3}r_{B}^{3}A_{1}^{0}B_{1}^{0}}{3(R_{AB})^{3}} + \thinspace O((\frac{1}{R_{AB}})^{5})
\end{align}

Now looking at the Jackson formula, we note that the distance between the two dipoles will be $R_{AB}$. Since the dipoles could either be aligned in the same direction or in opposing directions, there is a plus or minus factor in front of the formula:
\begin{equation} E^{int}_{d-d}=\frac{\pm({p}_{1}{p}_{2}-3{p}_{1}{p}_{2})}{R^{3}}=\frac{\pm2{p}_{1}{p}_{2}}{R^{3}}=\frac{\pm 2(\sqrt{\frac{4\thinspace \pi}{3}}r_{A}^{3}A_{1}^{0})(\sqrt{\frac{4\thinspace \pi}{3}}r_{B}^{3}B_{1}^{0})}{(R_{AB})^{3}}=\frac{\pm 8\thinspace \pi r_{A}^{3}r_{B}^{3}A_{1}^{0}B_{1}^{0}}{3(R_{AB})^{3}} \end{equation}
which agrees with our lowest order term from the dipole-dipole interaction energy expansion.\\
\indent We can also compare a numerical result from our calculations with an analytical solution done by Maxwell\cite[p.275]{Maxwell}. Maxwell finds that two touching spheres of radius $r$ held at potential $V=1$ statvolt each have a charge of $Q=r \log(2)$. Taking $r=1$ cm and using the equations for the energy stored in a capacitor we find:
\begin{equation} C=\frac{Q_{total}}{V}=\frac{2\log(2)}{1}=2\log(2) \end{equation}
\begin{equation} E_{cap}=\frac{1}{2}\frac{Q_{total}^{2}}{C} = \frac{1}{2}\frac{2^{2}}{2\log(2)}=\frac{1}{\log(2)} \approx 1.442695 \quad \text{ergs}\end{equation}
where $E_{cap}$ is the energy stored in the capacitor. This is extremely close to our energy values for the $R_{AB}=2$ limiting case. In addition, our formula converges rather quickly to the correct value.
\begin{table}
\caption{\label{tab:table1} Comparison of Calculated Energies with Analytic Maxwell Energies}
\begin{ruledtabular}
\begin{tabular}{ccc}
\textrm{Highest Order Term}&
\textrm{Energy (ergs)}&
\textrm{Percent Difference}\\
\colrule
Dipole-Dipole & 1.450672 & 0.553 \% \\
Quadrupole-Quadrupole & 1.444552 & 0.128 \% \\
Octupole-Octupole & 1.443789 & 0.076 \% \\
\end{tabular}
\end{ruledtabular}
\end{table}
To calculate the energies reported above, we explicitly performed the integrations from our final energy equation using Mathematica. We then minimized the energy equation utilizing the matrix formalism detailed in the next section. The Gegenbauer expansion of the denominator, $\mathcal{R}$, was taken out to the sixth Gegenbauer polynomial. This was the point at which we started getting marginal returns in the percent difference between our calculated energy and Maxwell's figures. The second Gegenbauer expansion used in the expansion of $\theta^{'}$ was taken out to the first term. This provided improvement over the small angle approximation and left the integrals still analytically computable by Mathematica.

\subsection{Generalizing for more spheres}
These formulas are generalized for any number of spheres by writing the multipoles as a vector and the individual terms of the energy as the entries of a matrix. This gives us the following matrix equation for the energy between any number of spheres:
\begin{equation}
\label{eq:matrix}
E = \frac{1}{2} \textbf{M}^{\dagger} \mathbb{K} \textbf{M} \; , \mathrm{with} \; \textbf{M}= \begin{pmatrix} A_{0}^{0} & A_{1}^{0}  \hdots & B_{0}^{0}  \hdots & C_{0}^{0} \hdots \end{pmatrix}^{T}\end{equation}
where $\textbf{M}$ is a vector containing all the multipoles for all the spheres to a particular order and $\mathbb{K} $ is the energy matrix that contains all the coefficients for the different multipole interactions. The diagonal terms of $\mathbb{K}$ are the self-energy terms, while generally, $\mathbb{K}_{n,m}$ is the interaction between the $\textbf{M}_{n}$ and the $\textbf{M}_{m}$ multipoles. The left section of $\mathbb{K}$ for a three sphere calculation up to dipoles is produced in the appendix. The minimum of equation \ref{eq:matrix} occurs in the trivial case when all the multipoles are equal to zero. To examine the non-trivial cases we need to find the constrained minimum by minimizing the function:
\begin{equation} \label{eq:langmatrix}
\mathcal{F}=\frac{1}{2} \textbf{M}^{\dagger} \mathbb{K} \textbf{M} -\sum \limits_{i} \lambda_{i} (\vec{c}_{i} \cdot \textbf{M}-Q_{i})
 \end{equation}
where $\lambda_{i}$ are the undetermined multipliers for the constraints. The term $\vec{c}_i$ is a dummy variable that uses the dot product $\vec{c}_{i} \cdot \textbf{M}$ to select out a specific multipole to constrain. The variable $Q_{i}$ is the constrained value for that specific multipole. For example, to constrain the monopole term on sphere A to be one, the second term of the equation would take the following form:
\begin{equation} \lambda (\vec{c} \cdot \textbf{M}-Q) \; = \; [\begin{pmatrix} 1 \\ 0 \\ \vdots \end{pmatrix} \cdot \begin{pmatrix} A_{0}^{0} \\ A_{1}^{0}\\ \vdots \end{pmatrix}-1]= \lambda (A_{0}^{0} -1)
\end{equation}

While we can arbitrarily constrain any multipole, our calculations only constrain the monopole terms since they correspond to the total charge on each sphere. For a configuration with three spheres, there are three different constraint equations; one for the monopole term on each sphere.

 In order to minimize $\mathcal{F}$, we take the derivative of equation \ref{eq:langmatrix} and set it equal to zero. The derivative is easy to calculate since $\mathbb{K}$ is a Hermitian matrix, and this yields the multipoles, $\textbf{M}$, that minimize the constrained equation:
\begin{align*} \nabla \mathcal{F} & =\mathbb{K}\textbf{M}-\sum \limits_{i} \lambda_{i} \vec{c}_{i}=0 \\
\Rightarrow \; & \; \textbf{M} =\mathbb{K}^{-1}\sum \limits_{i} \lambda_{i} \vec{c}_{i} \end{align*}
\indent By using this matrix formalism and solving for the unconstrained multipliers we can find the multipoles that minimize the electrostatic energy for any given number of charged conducting spheres.

\section{Data - Energy of Specific Configurations}
Here we present the configuration energy and the work required to remove a bounded sphere for various configurations. The configuration energy includes both the interaction and self energies of all the spheres except the monopole self energy. Since we constrain the monopole terms to constant values, we omit them from the energy data. The work required to remove a sphere is calculated by finding the difference in configuration energy between the original configuration and the configuration with one sphere removed. In the case of two spheres, the work needed to remove a sphere is the same as the configuration energy. In the case of three spheres, we subtract the total configuration energy from the energy of the two-sphere system that results after we remove a sphere. For asymmetrical configurations, we compare the work required to remove each of the different spheres and report the minimum.

The spheres in the figures are shaded with their charge density. The coloring goes from  blue to red as the charge density goes from negative to positive (color online, dark to light in grayscale). In addition there are lines drawn around the z-axis of each sphere to help qualitatively show the orientation of the charge on the spheres. The figures are meant to show alignment of the various higher-order multipoles relative to the other spheres in its own figure. The shading scale was varied between different figures to provide enough contrast within a figure.

In constructing these configurations, we label spheres with letters (A,B,C...) usually placing sphere A at the origin as a convention. Since the equations are rotationally invariant usually only one specific configuration is present (i.e. no more than one equilateral triangle with one charged sphere will be looked at since all three possibilities are equivalent).

For the calculations, we used the same parameters discussed in section \ref{sec:check} when we compared our energy values with Maxwell's calculations. Charged spheres had their monopole constrained so that the sphere has one statcoulomb of charge. All spheres have radius of 1 cm, and all energies are reported in ergs (Gaussian or cgs units).
\subsection{Two Sphere Configurations}
\begin{table}
\caption{\label{tab:table2} Two spheres \footnote{The energies differ by 1 compared with the energies cited in the Maxwell's calculation since we do not include the self-energy due to the constrained monopole.}}
\subtable[Both Charged]{
\begin{ruledtabular}
\begin{tabular}{c | c}
\textbf{Highest Order Interaction} & \textbf{Energy} (ergs)\\
\hline
Octupole-Octupole & 0.443789 \\
Quadrupole-Quadrupole & 0.444552 \\
Dipole-Dipole & 0.450672 \\
\end{tabular}
\end{ruledtabular}}
\subtable[One charged]{
\begin{ruledtabular}
\begin{tabular}{c|c}
\textbf{Highest Order Interaction} & \textbf{Energy}  (ergs)\\
\hline
Octupole-Octupole & -0.0557567 \\
Quadrupole-Quadrupole & -0.048093 \\
Dipole-Dipole & -0.0336496 \\
\end{tabular}
\end{ruledtabular}}
\end{table}

\begin{figure}
\centering
\subfigure[Both charged]{
\includegraphics[width=2in] {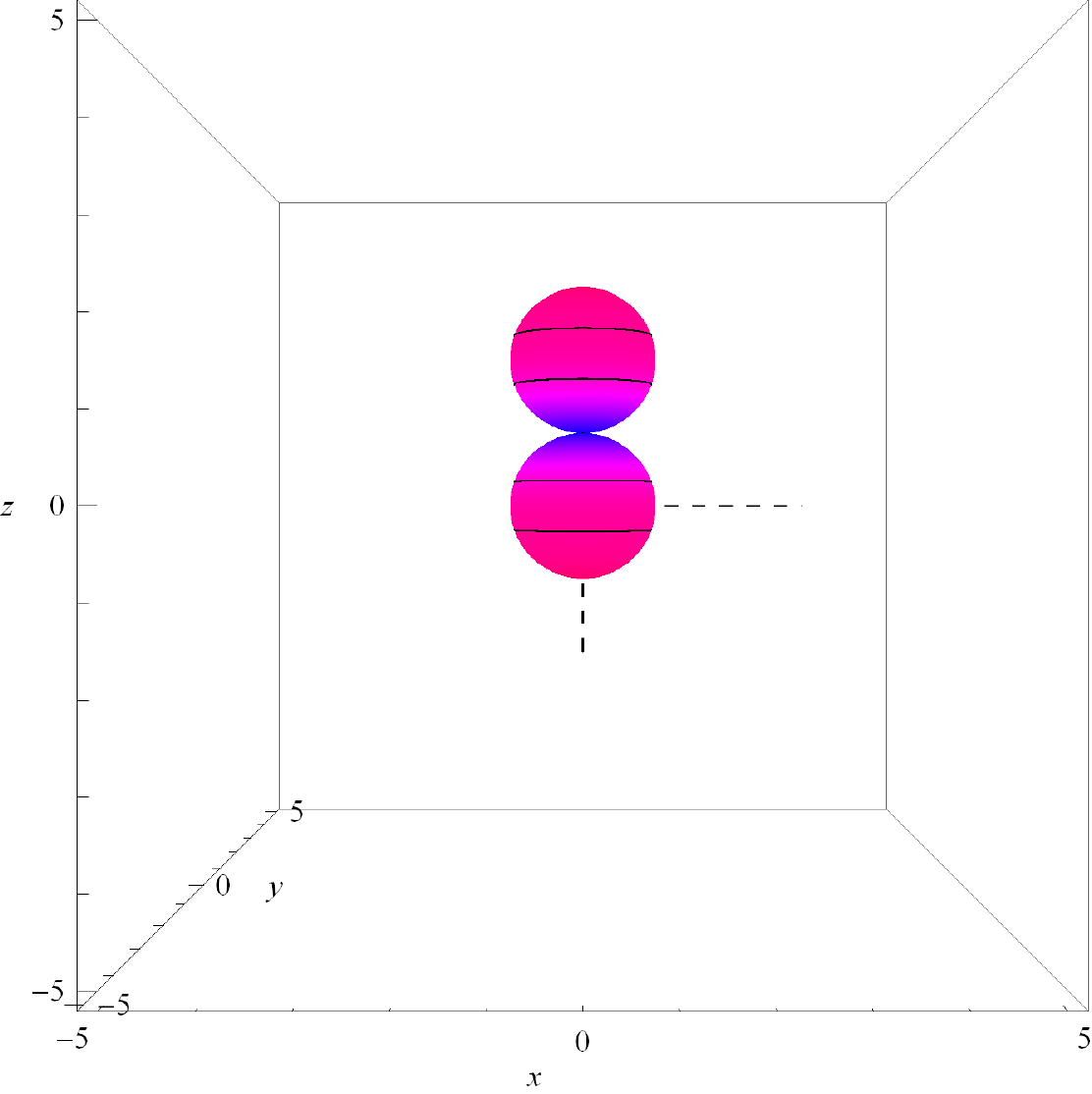}
\label{fig:subfig1.41}}
\subfigure[Bottom sphere charged]{
\includegraphics[width=2in] {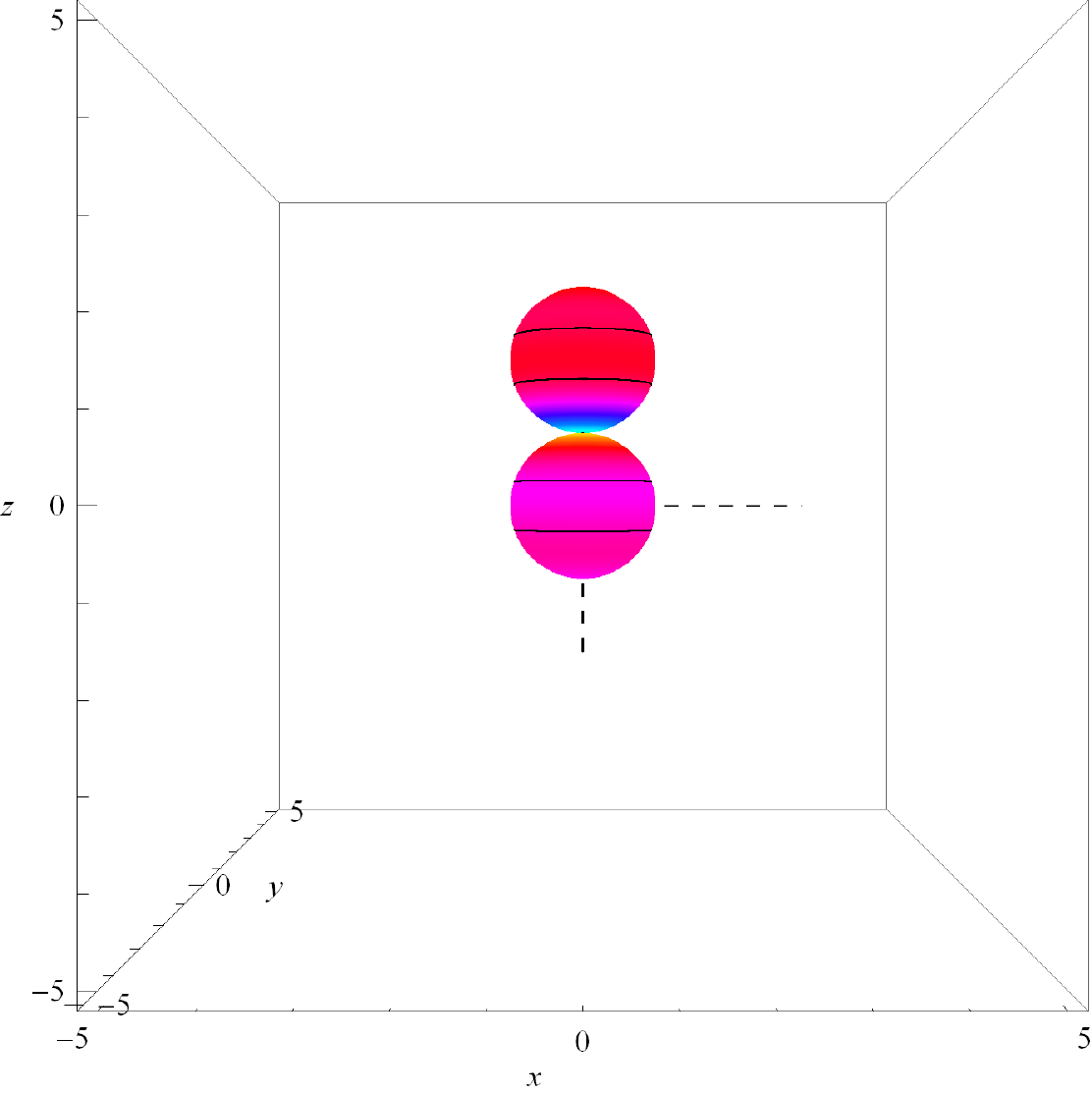}
\label{fig:subfig1.42}}
\caption{Comparison of the two configurations with the total charge distribution shaded on the surface.}
\label{fig:4}
\end{figure}
\begin{figure}
\centering
\subfigure[Both charged]{
\includegraphics[width=2in] {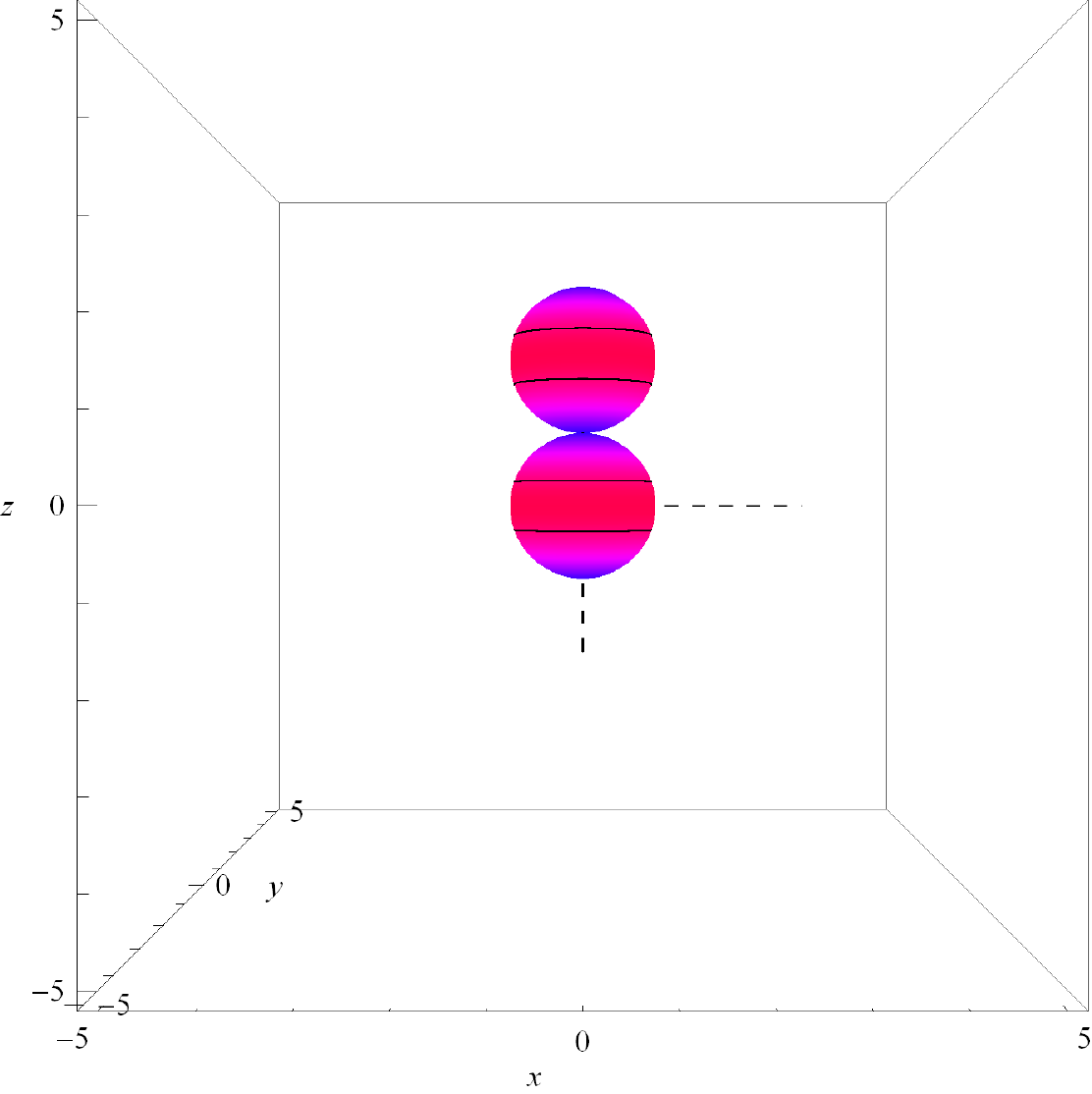}
\label{fig:subfig1.51}}
\subfigure[Bottom sphere charged]{
\includegraphics[width=2in] {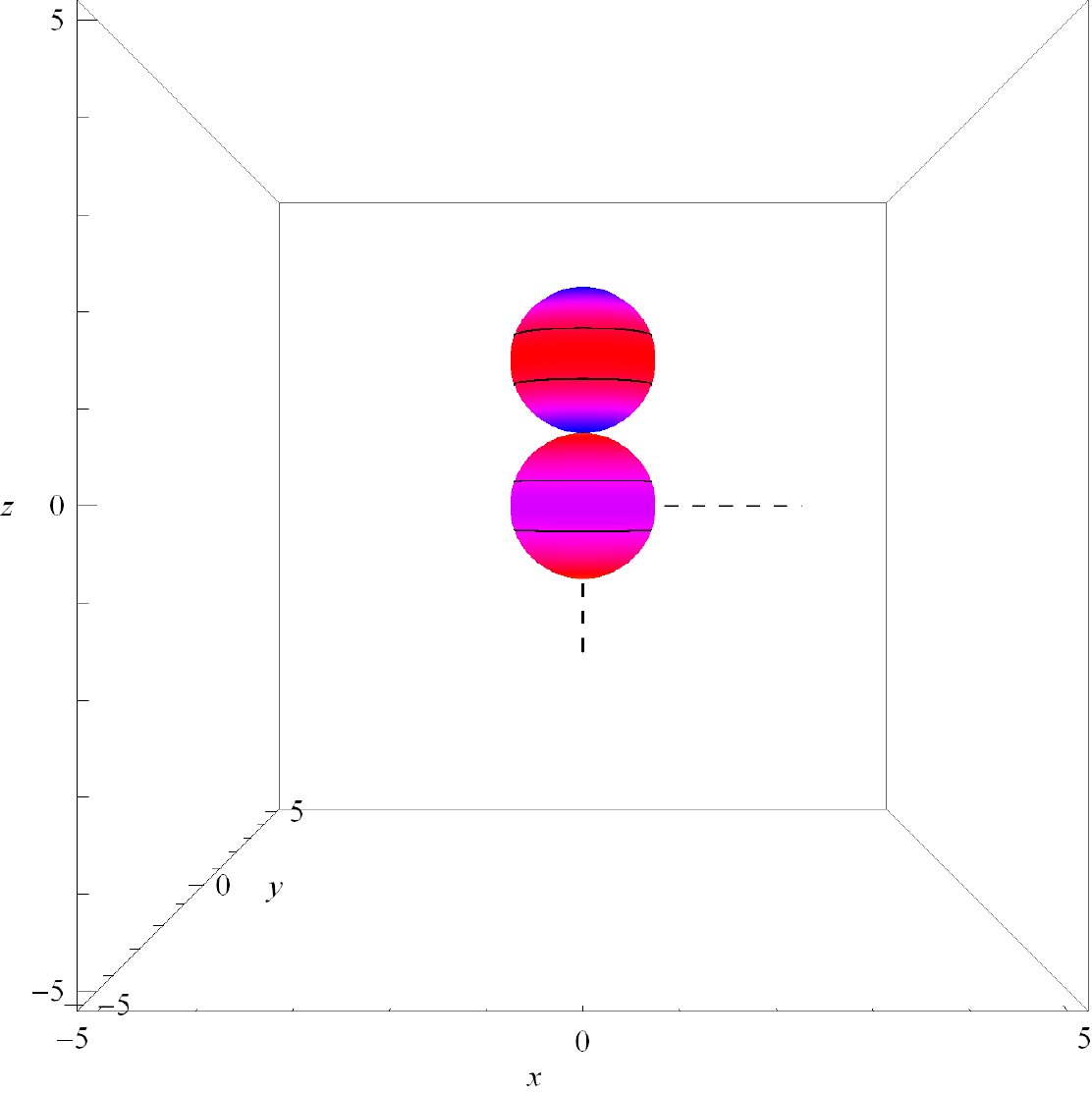}
\label{fig:subfig1.52}}
\caption{Comparison of the two configurations with only the quadrupole charge density shaded on the surface.}
\label{fig:5}
\end{figure}
\subsubsection{Discussion of Two Sphere Configurations}
In the case of two spheres, there are only two different configurations of interest. The first is when both spheres have equal charges. This induces multipoles equal in magnitude for both spheres and equal binding energies for each sphere. When only one sphere is charged, the higher order multipole terms have a much stronger role. Looking back at the explicit interaction energy expansion, the first three terms of the energy expansion are the monopole-monopole energy and the two monopole-dipole interactions. However, only one of those three terms is non-zero when one sphere is left uncharged. This results in the binding energy for the more weakly bound sphere (Sphere B) to increase 43\% when the expansion is extended from the dipole-dipole interaction up to the quadrupole-quadrupole interaction. If we take the calculation up to the octupole-octupole interaction, the binding energy of that sphere increases by another 15\%. This is compared to only about a 0.2\% difference between the quadrupole-quadrupole and octupole-octupole calculations in the configuration with both spheres charged. If we had arbitrarily cut off the calculation at the dipole-dipole term of the configuration with only one sphere charged, we would have seen a binding energy for sphere B that is about one-third of the strength it actually is. This shows evidence for why higher ordered terms cannot be arbitrarily ignored in these energy calculations. Also, since the configuration energy relies heavily on higher order multipoles, the energy values do not converge quickly when only one sphere is charged.
\subsection{Three Sphere Case}
\begin{sidewaystable}
\begin{figure}[H]
\centering
\includegraphics [width=20cm] {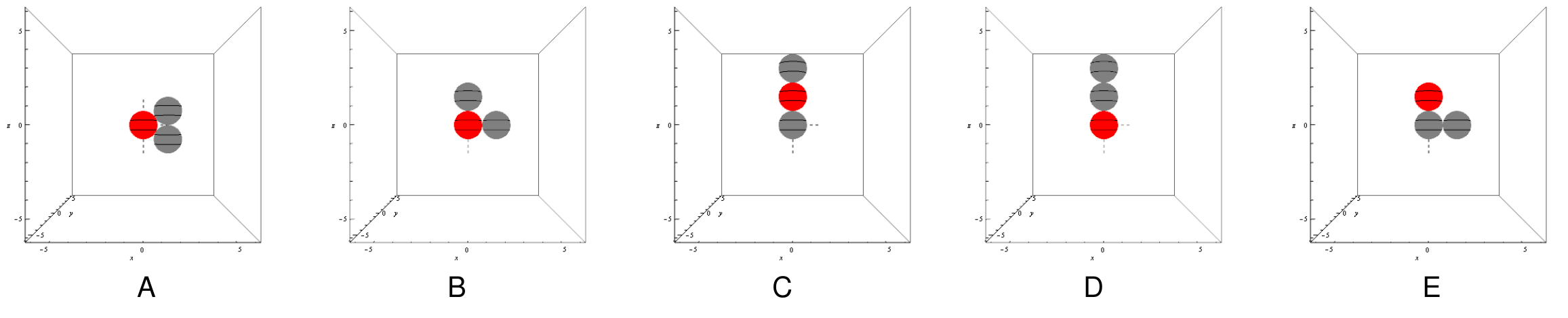}
\caption{Diagram of each configuration. Red spheres have positive charge of one unit, while the dark gray spheres are uncharged (color online).}
\label{fig:}
\end{figure}
\end{sidewaystable}
\begin{table*}
\caption{\label{tab:table3} Three spheres}
\subtable[Total configuration energies vs. Highest order interaction (energy in ergs)]{
\begin{ruledtabular}
\begin{tabular}{c|ccccc}
& A & B & C & D & E \\
\hline
Octupole-Octupole & -0.111602 & -0.0999026 & -0.09723 & -0.0702905 & -0.0690104 \\
Quadrupole-Quadrupole & -0.0965803& -0.0871968 & -0.0889388 & -0.0570036& -0.0602305 \\
Dipole-Dipole & -0.0596624 & -0.0627721 & -0.0605811 & -0.0438303 & -0.0400716
\end{tabular}
\end{ruledtabular}}
\subtable[Work to remove weakest bound sphere to infinity vs. Highest order interaction  (energy in ergs)]{
\begin{ruledtabular}
\begin{tabular}{c|ccccc}
& A & B & C & D & E \\
\hline
Octupole-Octupole & 0.055845 & 0.0441459 & 0.0414733 & 0.0145338 & 0.0132537 \\
Quadrupole-Quadrupole & 0.0484873 & 0.039103 & 0.0408458 & 0.00891062& 0.0121374 \\
Dipole-Dipole &0.0260128 & 0.0291225 & 0.0269315 & 0.0101807 & 0.00642202
\end{tabular}
\end{ruledtabular}}
\end{table*}

\begin{figure}[H]
\centering
\subfigure[Equilateral triangle (config. A)]{
\includegraphics[width=2in] {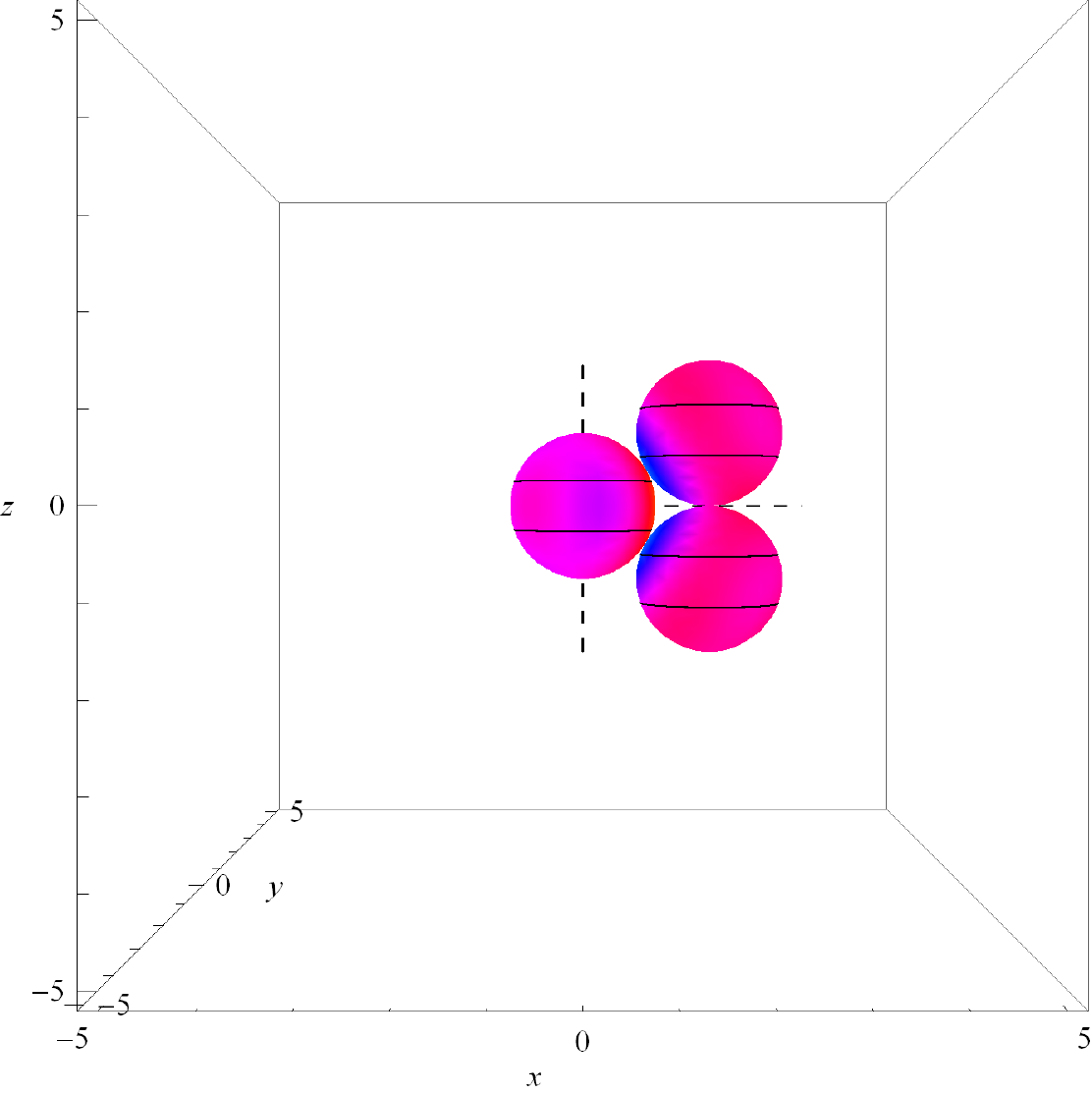}
\label{fig:subfig6.1}}
\subfigure[Right triangle, charged apex (config. B)]{
\includegraphics[width=2in] {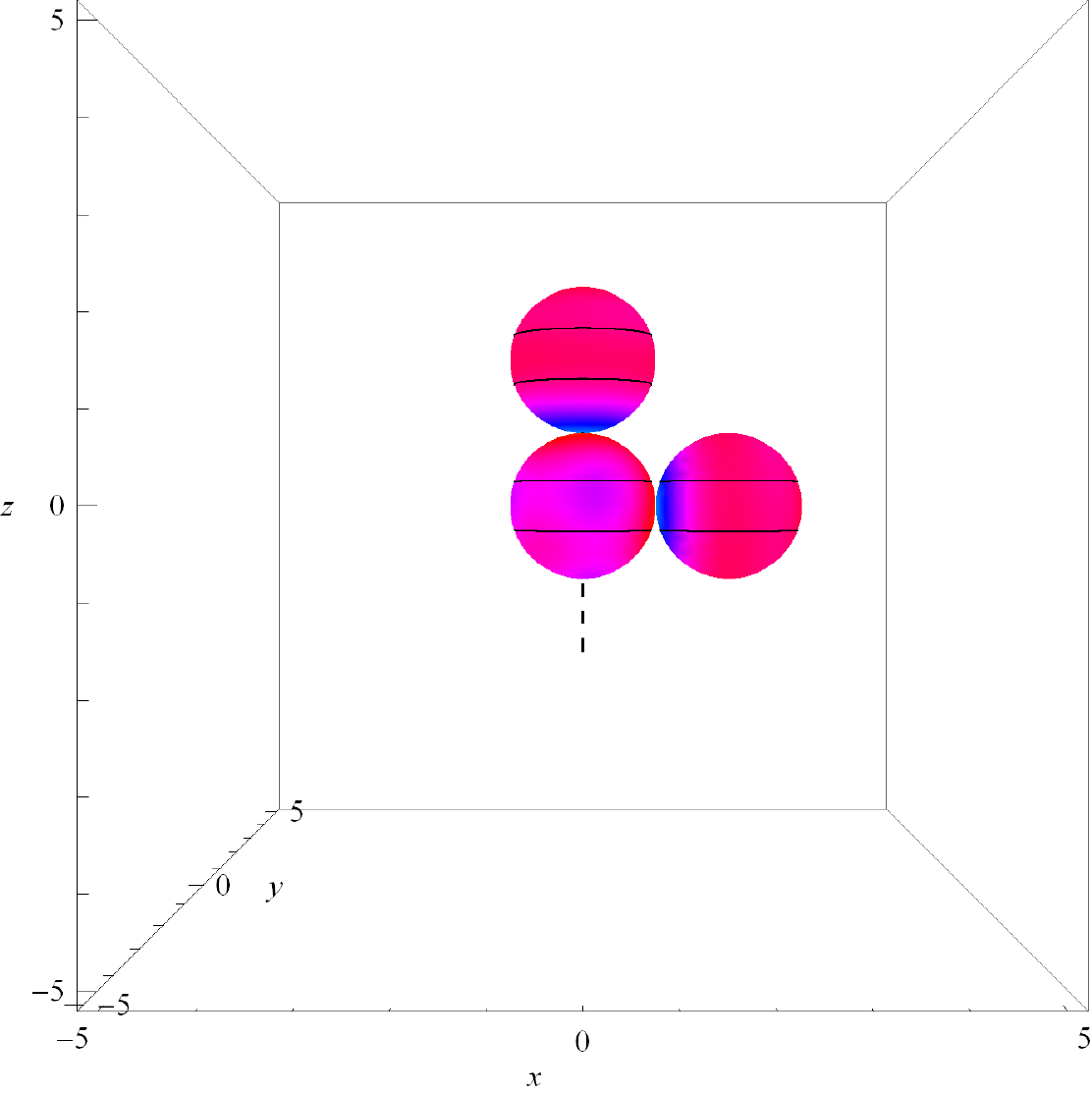}
\label{fig:subfig6.2}}
\caption{Comparison of two configurations with the total charge configuration shaded on the surface}
\label{fig:6}
\end{figure}
\begin{figure}[H]
\centering
\subfigure[Equilateral triangle (config. A)]{
\includegraphics[width=2in] {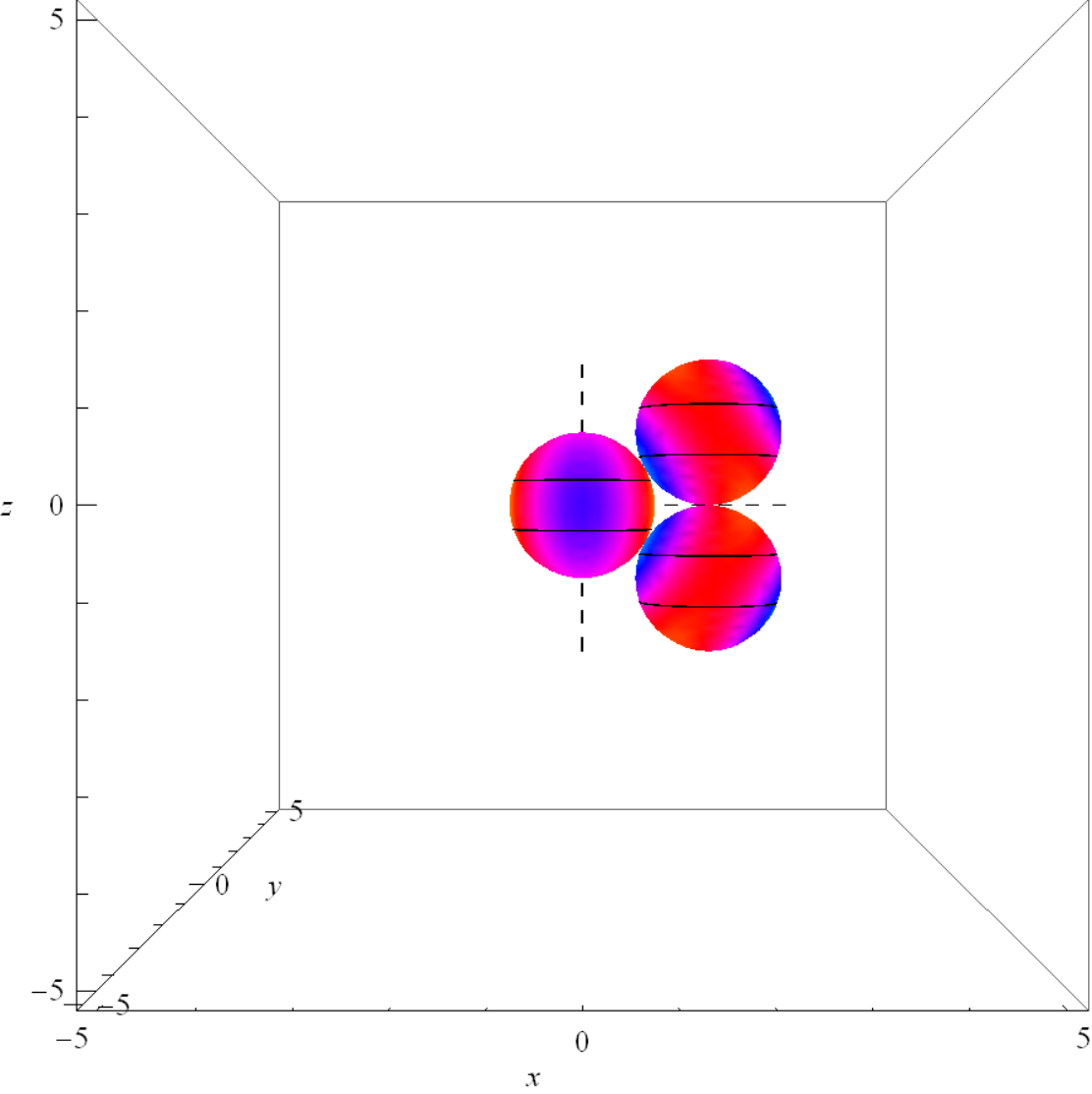}
\label{fig:subfig7.1}}
\subfigure[Right triangle, charged apex (config. B)]{
\includegraphics[width=2in] {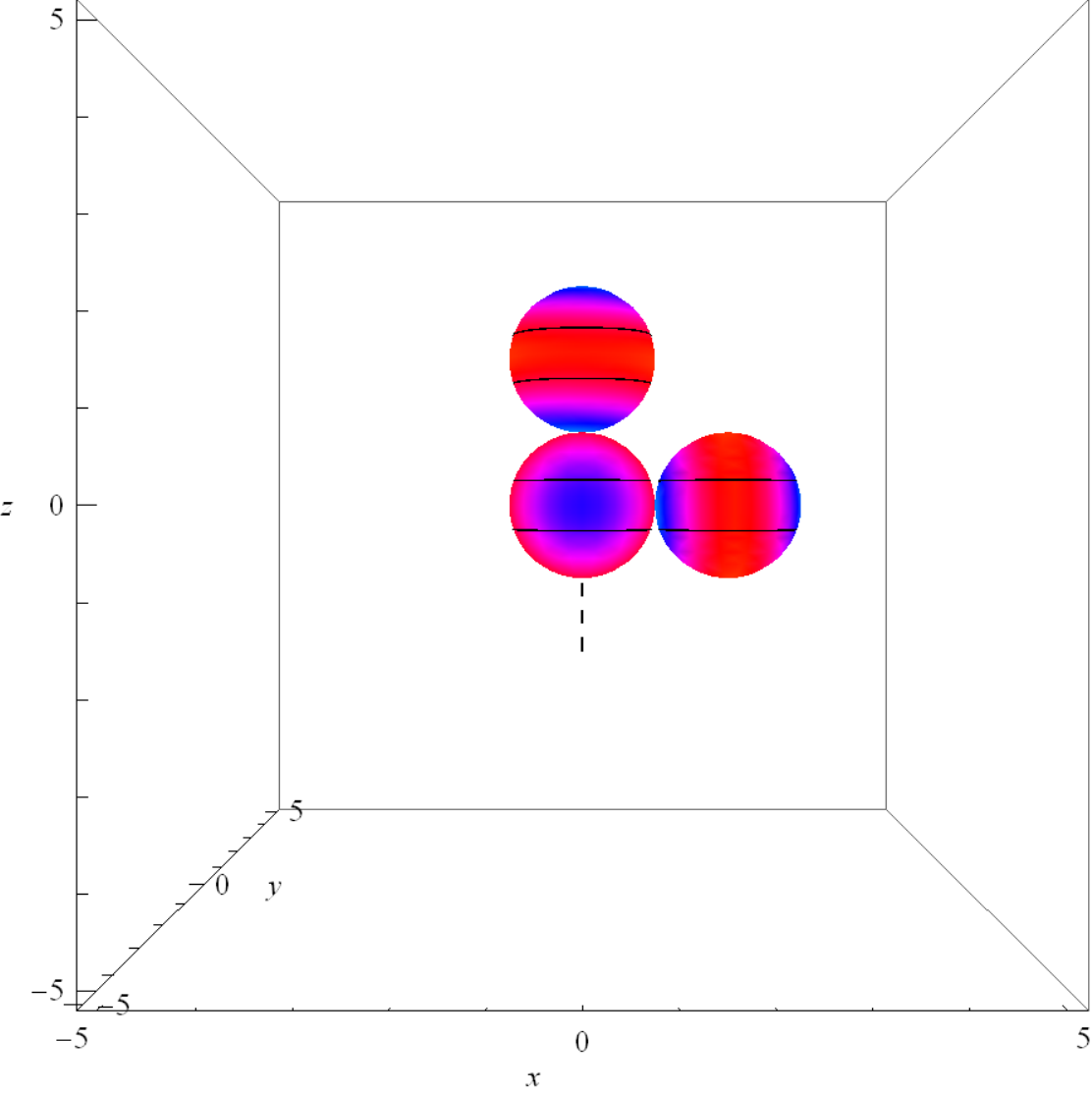}
\label{fig:subfig7.2}}
\caption{Comparison of the two configurations with only the quadrupole charge densities shaded on the surface}
\label{fig:7}
\end{figure}

For the three sphere case, the minimal energy configuration is the equilateral triangle (configuration A). This is followed by the isosceles triangle (config. B) and the middle-charged line of spheres (config. C). As more multipoles are added, the minimal energy configuration changes. When only dipoles are taken into account, both the isosceles triangle and the middle-charged line of spheres are favored over the equilateral triangle. This is due to minimal dipole-dipole interaction between the two uncharged spheres in each configuration.  The isosceles triangle configuration induces mutually orthogonal dipoles on the two uncharged spheres, reducing the energetic cost of the configuration. In the middle-charged line of spheres, the monopole on the middle sphere dominates the energy equation through the monopole-dipole terms. The antiparallel dipoles that are induced on the two uncharged spheres have only a slight effect on the configuration energy. However, this effect is strong enough to give the configuration a slightly higher energy than the isosceles triangle.

Once we add quadrupoles, the dipole advantages discussed above are much less influential, causing the equilateral triangle becomes the minimal energy configuration. Since all the dipoles in the equilateral triangle line up to the same point in space, there is much more repulsive interaction between the dipoles than in the isosceles triangle configuration. This caused it to have a misleadingly higher configuration energy when only dipoles are considered. With the quadrupoles added, the equilateral triangle experiences the largest increase in binding energy, making it the lowest energy configuration. The symmetry of this configuration causes the energy to be more dependent on quadrupole terms than the isosceles triangle or the middle-charged line of spheres. This shows that some configurations show non-regular convergence when different multipoles are added to the calculation. In particular, the work to remove a sphere from configuration D is not even monotonically decreasing as we add higher order terms, demonstrating the significant effect that the geometry of the configuration has on the binding energies. 

In addition, configurations with uncharged spheres equidistant from the charged sphere have stronger binding energies. The line with a charge on the end (config. D) and the isosceles triangle with a charge on one of the legs (config E) are the most weakly bound configurations. Both of these configurations are asymmetrical in the sense that the two uncharged spheres are not equidistant from the charged sphere. Judging from the charge distribution diagrams of these configurations, it seems that the asymmetric configurations restrict the polarization of one of the uncharged spheres. This prevents the configuration energy from lowering as far as the other three configurations that all feature the uncharged spheres equidistant from the charged sphere. 

We also see that the energy convergence of the equilateral triangle's configuration energy is similar to the the two sphere case. The binding energy increases by 86\% when the calculation order is taken from dipoles up to quadrupoles and a further 16\% when octupole interactions are added in, comparable to the convergence seen in the two sphere case. However, there is variability in energy convergence for different configurations. For example, the binding energy for configurations C and D only increase by 1.5\% and 8\% respectively when the calculation is extended from quadrupoles to octupoles.

\subsection{Four Sphere Configurations}
\begin{sidewaystable}
\begin{figure}[H]
\includegraphics [width=20cm] {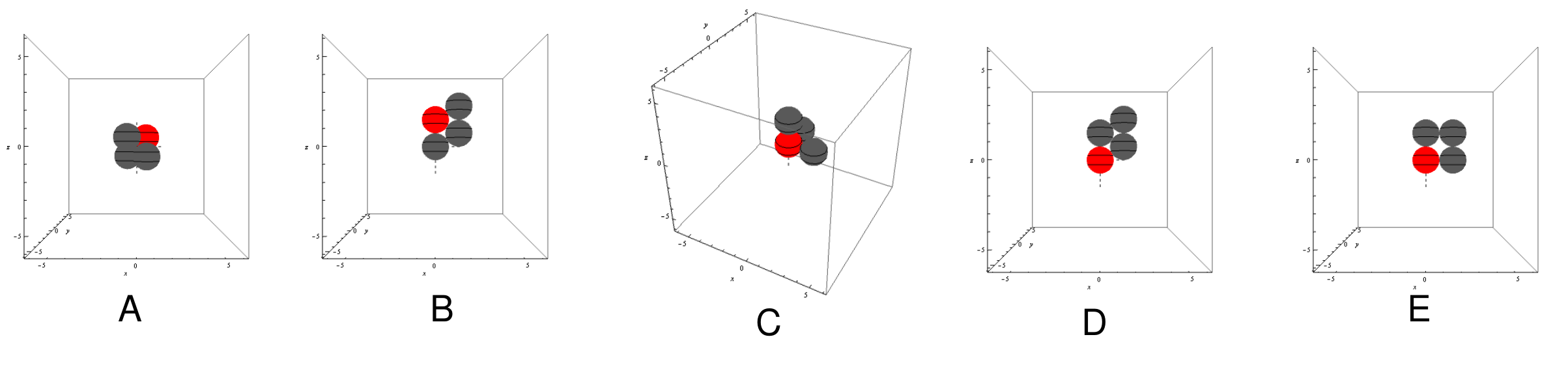}
\caption{Diagram of each configuration. Red spheres have positive charge of one unit, while the dark gray spheres are uncharged (color online).}
\label{fig:config4}
\end{figure}
\end{sidewaystable}
\begin{table*}
\caption{\label{tab:table4} Four spheres}
\subtable[Total configuration energies vs. Highest order interaction  (energy in ergs)]{
\begin{ruledtabular}
\begin{tabular}{c|ccccc}
& A & B & C & D & E \\
\hline
Octupole-Octupole & -0.169347 & -0.156567 &-0.135689& -0.129386 & -0.112597 \\
Quadrupole-Quadrupole & -0.148192 & -0.13541 &-0.119614& -0.11312&-0.099499 \\
Dipole-Dipole & -0.0804022 & -0.0822249 &-0.0882232& -0.068025 &-0.0672607
\end{tabular}
\end{ruledtabular}}
\subtable[Work to remove weakest bound sphere to infinity vs. Highest order interaction  (energy in ergs)]{
\begin{ruledtabular}
\begin{tabular}{c|ccccc}
& A & B & C & D & E \\
\hline
Octupole-Octupole & 0.0577453 & 0.0449653 &0.0357864& 0.0177843& 0.0126944 \\
Quadrupole-Quadrupole & 0.0516117 & 0.0388297 & 0.0324172 & 0.0165397& 0.0123022 \\
Dipole-Dipole & 0.0207398 & 0.0225625 &0.0254511 &0.00836263 & 0.00448864
\end{tabular}
\end{ruledtabular}}
\end{table*}

\begin{figure}[H]
\centering
\subfigure[Blunt end charged (config B)]{
\includegraphics[width=2in] {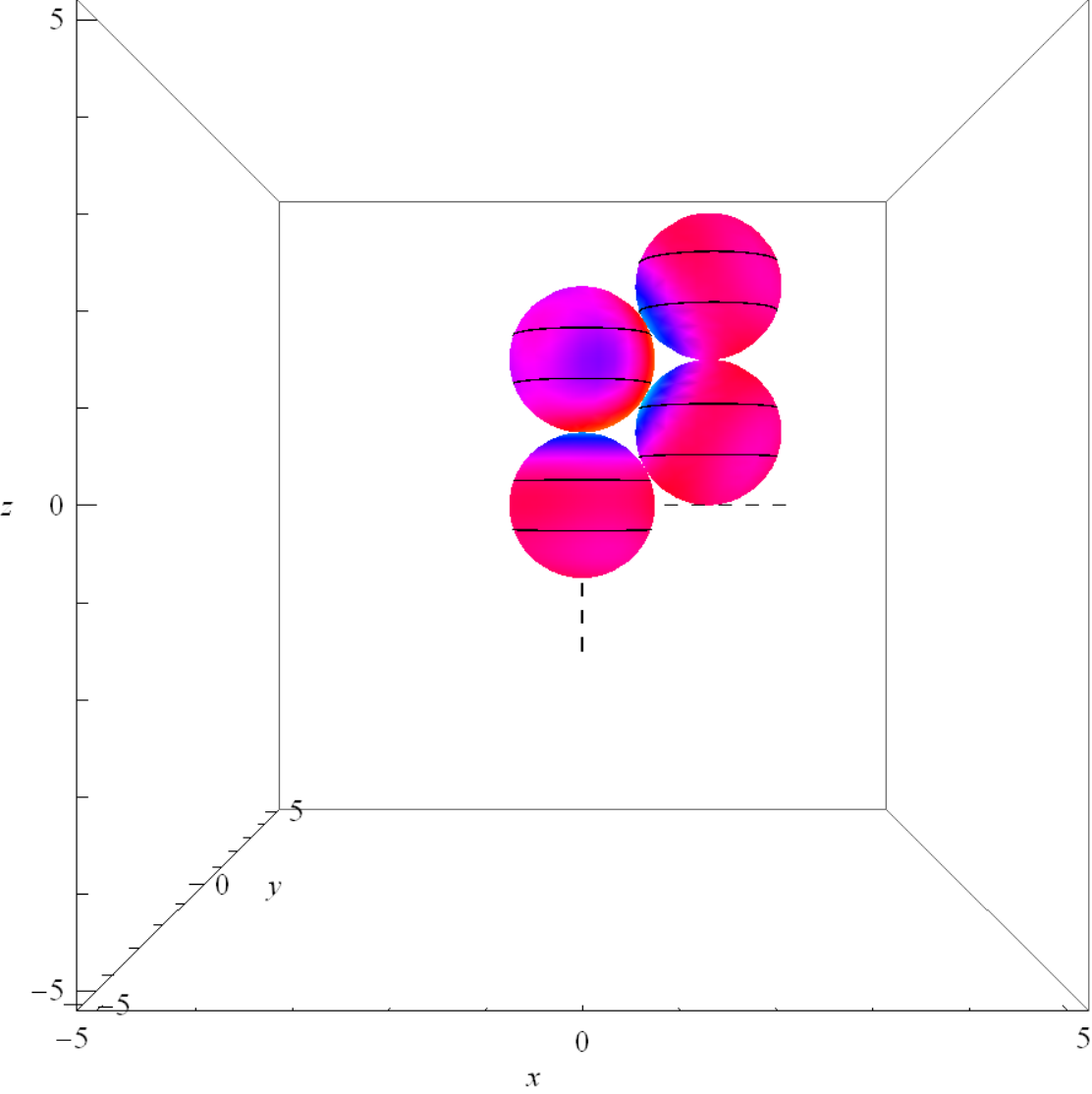}
\label{fig:subfig8.1}}
\subfigure[Sharp end charged (config D)]{
\includegraphics[width=2in] {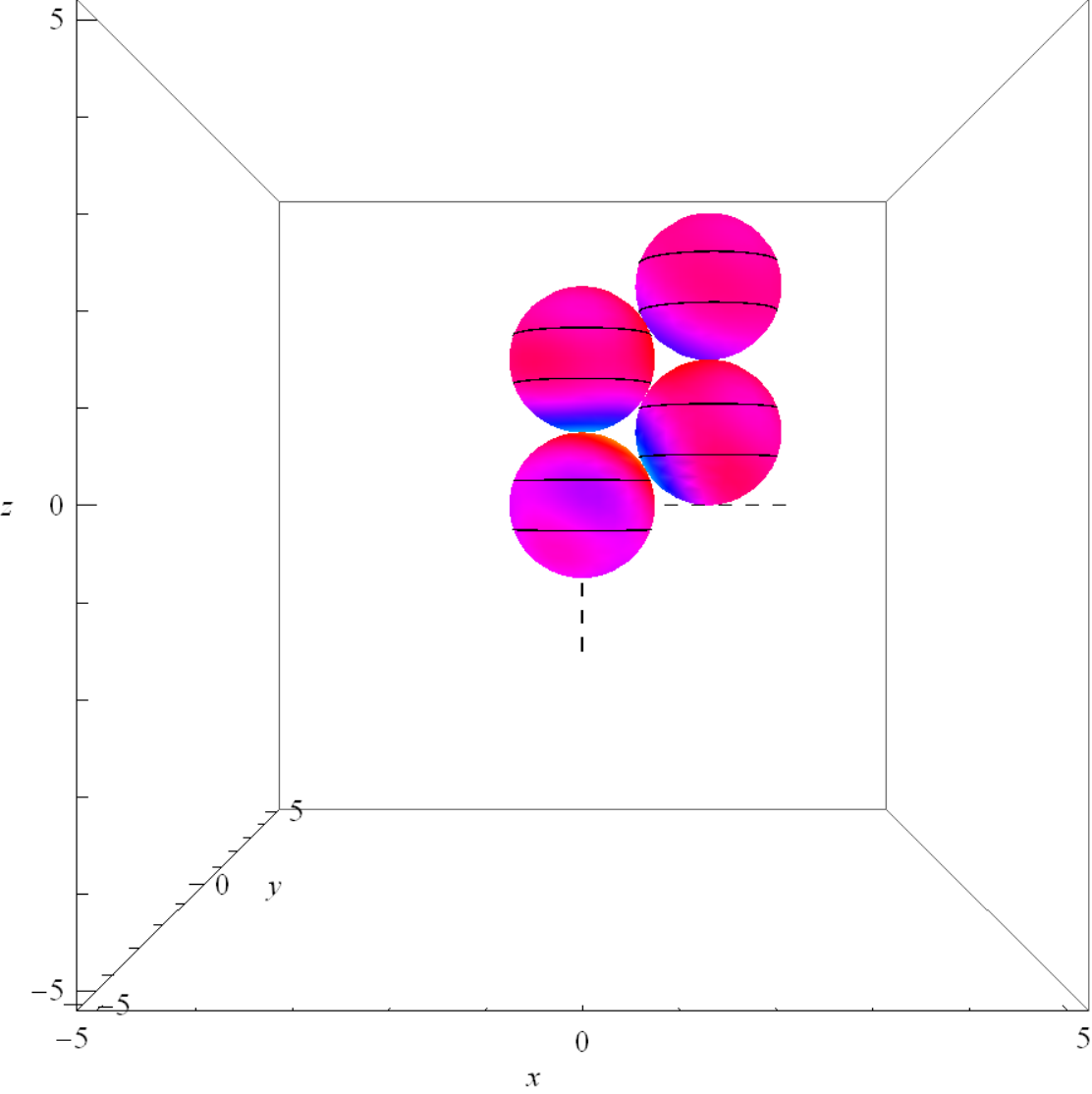}
\label{fig:subfig8.2}}
\caption{Comparison of two rhombus configurations with the total charge configuration shaded on the surface}
\label{fig:8}
\end{figure}
\begin{figure}[H]
\centering
\subfigure[Blunt end charged (config B)]{
\includegraphics[width=2in] {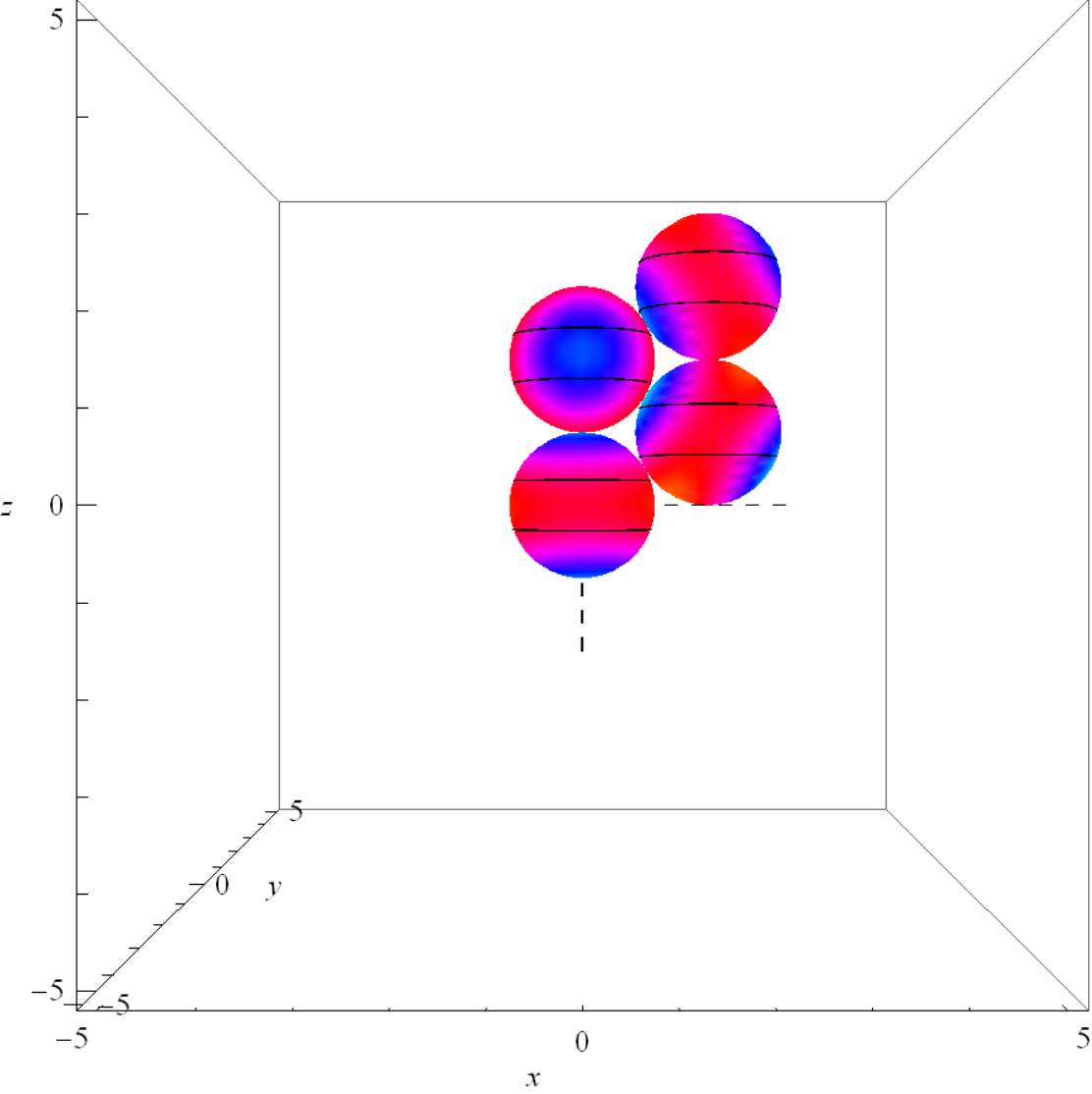}
\label{fig:subfig9.1}}
\subfigure[Sharp end charged (config D)]{
\includegraphics[width=2in] {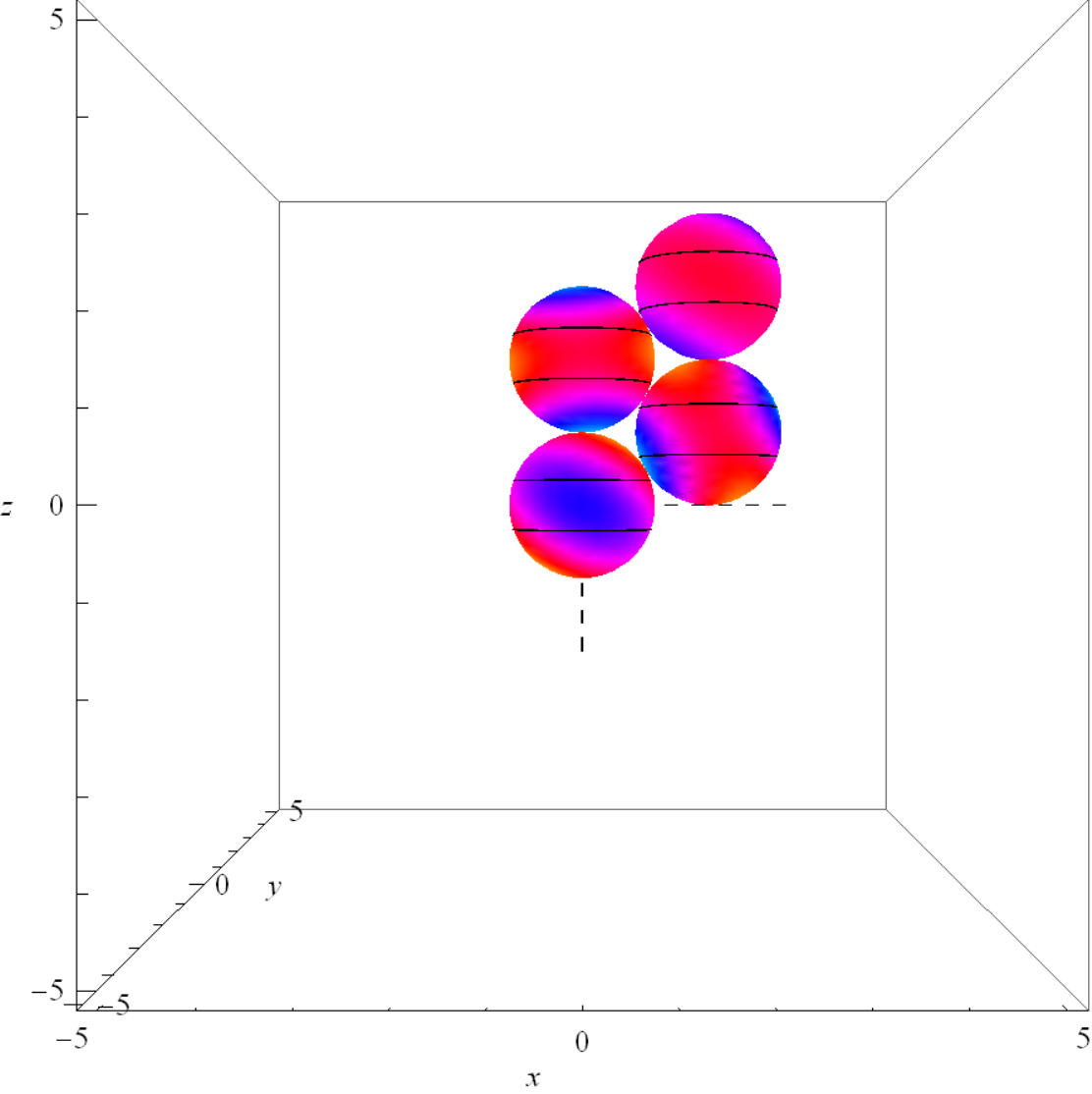}
\label{fig:subfig.9.2}}
\caption{Comparison of the two configurations with only the quadrupole charge densities shaded on the surface}
\label{fig:9}
\end{figure}

For configurations with four spheres, the minimal energy configuration is the regular tetrahedron (config. A). Much like in the three-sphere case, the minimal energy configuration changes as additional multipole terms are added to the calculation. When only dipoles are considered, the minimal energy configuration is the non-regular tetrahedron (config. C). Since most of the faces of this tetrahedron are isosceles triangles, the behavior of this configuration is similar to the isosceles triangle that we analyzed in the three-sphere case. As before, the configurations that induce dipoles in mutually orthogonal direction will appear to have an energetic advantage when the calculation only includes dipole terms.

The regular tetrahedron (config. A) mimics the behavior of the equilateral triangle, with all of the spheres' induced dipoles focusing on a common point. This causes strong interactions between the induced dipoles and makes configuration A look less favorable when only including dipole interactions in the calculation. Thus, when only dipoles are included in the calculations, there is an intrinsic bias for configurations that feature mutually orthogonal dipoles. This bias, caused by arbitrarily cutting off the calculation at the dipole-dipole interaction, results in incorrect conclusions for the minimal energy configuration.

Once quadrupoles are added, the configurations take the same order as when the calculation is carried out to octupoles. The regular tetrahedron (config. A) is the preferred configuration with the charged blunt-end rhombus as the next most favored. Note the significant energy difference between the two rhombus configurations (configs. B and D). When the blunt-end particle is charged in a rhombus, the charged sphere is in contact with all of the other three spheres in the configuration causing significant polarizations in the three uncharged sphere. However, in the sharp-end charged rhombus (config. D), the charged sphere is only in contact with two other spheres. This causes a much smaller polarizing effect on the third sphere and makes the configuration less energetically favorable when compared with the blunt-end charged rhombus (config B).
\section{Conclusion}
Here we presented a new method utilizing multipole expansions to calculate the energy of charged conducting sphere configurations. This method can be used for any number of spheres in an arbitrary configuration, and it is easily expanded to any multipole order. 

As seen from the configurations we considered, cutting off the energy expansion at arbitrary multipole orders (like at the dipole-dipole or quadrupole-quadrupole level) gives an incorrect picture of the energetically preferred configurations. In addition, there are intrinsic biases towards particular configurations. Configurations that induce mutually orthogonal dipoles, were favored when we cut off the energy expansion at the dipole-dipole interaction.

In this study, we truncated the energy expansion at the octupole-octupole level. In the case of two charged spheres, this was sufficient since the energy converged quickly; the energy only increased 0.2\% when the calculation was taken from quadrupoles up to octupoles. However, this is not the case when we move on to more general configurations with multiple spheres. In general, the more spheres in a configuration, the more the energy calculations rely on higher order multipoles, and the less quickly the energy converges.  The calculations here were also subject to an arbitrary cut-off for interactions at the octupole-octupole level. Given more computing power or patience, the calculations could be extended even further.

We can extend this study by analyzing further how different multipoles interact in various configurations. For example, a family of configurations could be found where the configuration energy only slightly changes with the addition of quadrupoles. This implies that the specific geometry of the configuration reduces or neutralizes the quadrupole-quadrupole interactions. By identifying similar configurations,  more complex tailor-made materials can be created that utilize configuration geometry to reduce individual inter-particle interactions.

This paper was the product of an independent research project done by Alex Moore under the supervision of Thomas Witten while at the University of Chicago. In addition Nathan Krapf, also at the University of Chicago, helped to extensively revise this paper. Alex would like to thank both Prof. Witten and Nathan Krapf for their patience and guidance while working on the project, as well as Dmitri Talapin for his support and for providing the initial motivation for this study.
\appendix
\section{Rotating the Multipoles is Equivalent to Rotating the Spherical Harmonics}
\label{app:linear}
\begin{proof}
The charge distribution on a sphere and the energy between two spheres are invariant with respect to any choice of coordinates. Therefore, given two systems of coordinates, $\chi=\{\theta,\phi\}$ and $\hat{\chi}=\{\hat{\theta},\hat{\phi} \}$, a multipole in the $\chi$ frame can be expanded in terms of the multipoles in the $\hat{\chi}$ frame for any point $\vec{x}=\{\theta,\phi\},\{\hat{\theta},\hat{\phi} \}$:
\begin{equation} A^{m}_{\ell}\thinspace Y^{m}_{\ell}(\theta,\phi)=\sum \limits_{M=-\ell}^{\ell} \hat{A}^{M}_{\ell}\thinspace Y^{M}_{\ell}(\hat{\theta},\hat{\phi})\end{equation}
Now we can rotate the $Y^{M}_{\ell}(\hat{\theta},\hat{\phi})$ into the $\chi$ frame by utilizing the identity \cite[p.218]{Landau}:
\begin{equation}
Y_{\ell}^{M} (\hat{\theta},\hat{\phi}) = \sum_{N=-\ell}^{\ell} [\mathcal{D}_{N,M}^{\ell}(\alpha,\beta,\gamma)\thinspace Y_{\ell}^{N} (\theta,\phi) ]\end{equation}
where $(\alpha,\beta,\gamma)$ are the Euler angles needed to rotate the $\hat{\chi}$ coordinate system to the $\chi$ coordinate system . Plugging this in:
\begin{equation} A^{m}_{\ell}\thinspace Y^{m}_{\ell}(\theta,\phi)=\sum \limits_{M=-\ell}^{\ell} \hat{A}^{M}_{\ell}(\sum_{N=-\ell}^{\ell} \mathcal{D}_{N,M}^{\ell}(\alpha,\beta,\gamma)\thinspace Y_{\ell}^{N} (\theta,\phi)) \end{equation}
\begin{equation} A^{m}_{\ell}\thinspace Y^{m}_{\ell}(\theta,\phi) =\sum_{N=-\ell}^{\ell}Y_{\ell}^{N} (\theta,\phi)(\sum \limits_{M=-\ell}^{\ell} \hat{A}^{M}_{\ell}\mathcal{D}_{N,M}^{\ell}(\alpha,\beta,\gamma)) \end{equation}
By the linear independence of the $Y^{m}_{\ell}$ 's, we must have $N=m$. This removes the summation over N and presents us with our final indentity:
\begin{equation} \Longrightarrow A^{m}_{\ell}=\sum \limits_{M=-\ell}^{\ell} \mathcal{D}_{m,M}^{\ell}(\alpha,\beta,\gamma)\hat{A}^{M}_{\ell} \end{equation}
If we are trying to rotate a conjugated multipole, we just use the rules of complex conjugation of sums and products to get the identity:
\begin{equation} \Longrightarrow A^{m*}_{\ell}=\sum \limits_{M=-\ell}^{\ell} (\mathcal{D}_{m,M}^{\ell}(\alpha,\beta,\gamma))^{*}\hat{A}^{M*}_{\ell} \end{equation}
\end{proof}

\begin{table*}
\caption{\label{tab:table3}Configuration Details. Location of charged sphere in bold.}
\subfigure[Three sphere configurations]{
\begin{ruledtabular}
\begin{tabular}{cccccc}
Configuration & Letter & Sphere A & Sphere B & Sphere C & Sphere D \\ \hline
 Equilateral Triangle & A & $\mathbf{(0,0,0)}$ & $(\sqrt{3},0,1)$ & $(\sqrt{3},0,-1)$ & - \\
 Isosceles - Apex Charged & B & $\mathbf{(0,0,0)}$ &  $(0,0,2)$ & $(2,0,0)$ & - \\
 Line - Middle Charged & C & $(0,0,0)$ & $\mathbf{(0,0,2)}$ & $(0,0,4)$ & - \\
 Line - End Charged & D & $\mathbf{(0,0,0)}$ & $(0,0,2)$ & $(0,0,4)$ & - \\
  Isosceles - Leg Charged & E & $(0,0,0)$ &  $(0,0,2)$ & $\mathbf{(2,0,0)}$ & -  \\
\end{tabular}
\end{ruledtabular}}
\subfigure[Four sphere configurations.]{
\begin{ruledtabular}
\begin{tabular}{cccccc}
Configuration & Letter & Sphere A & Sphere B & Sphere C & Sphere D \\ \hline
 Regular Tetrahedron & A & $\mathbf{(\frac{\sqrt{2}}{2},\frac{\sqrt{2}}{2},\frac{\sqrt{2}}{2})}$ & $(-\frac{\sqrt{2}}{2},-\frac{\sqrt{2}}{2},\frac{\sqrt{2}}{2})$ & $(-\frac{\sqrt{2}}{2},\frac{\sqrt{2}}{2},-\frac{\sqrt{2}}{2})$ & $(\frac{\sqrt{2}}{2},-\frac{\sqrt{2}}{2},-\frac{\sqrt{2}}{2})$ \\
Rhombus - Blunt Charged & B &  $(0,0,0)$ & $\mathbf{(0,0,2)}$ & $(\sqrt{3},0,1)$ & $(\sqrt{3},0,3)$\\
Non-Reg Tetrahedron & C & $\mathbf{(0,0,0)}$ & $(2,0,0)$ & $(0,2,0)$ & $(0,0,2)$ \\
Rhombus - Sharp Charged & D &  $\mathbf{(0,0,0)}$ & $(0,0,2)$ & $(\sqrt{3},0,1)$ & $(\sqrt{3},0,3)$\\
  Square & E & $\mathbf{(0,0,0)}$ & $(0,0,2)$ & $(2,0,0)$ & $(2,0,2)$ \\
\end{tabular}
\end{ruledtabular}}
\end{table*}

 \section{Explicit Matrix Equation}
 This section of the matrix $\mathbb{K}$ (defined in equation \ref{eq:matrix}) covers the projection of $\{A_{0}^{0},A_{1}^{-1},A_{1}^{0},A_{1}^{1}\}$ onto the  monopole and dipole terms from spheres A, B, and the monopole from sphere C (given by vector $\{A_{0}^{0},A_{1}^{-1},A_{1}^{0},A_{1}^{1},B_{0}^{0},B_{1}^{-1},B_{1}^{0},B_{1}^{1},C_{0}^{0}\}$). For simplicity, we assume the spheres are all the same size ($r_{A}=r_{B}=r_{C}=1$), and we take the limit $R_{AB}, R_{AC}, R_{BC} >>1$. This gives the first few rows and columns of $\mathbb{K}$ as:
$$\mathbb{K}=
\left(\begin{array}{cccc}
 4 \pi  & 0 & 0 & 0  \hdots \\
 0 & \frac{4 \pi }{3} & 0 & 0  \hdots \\
 0 & 0 & \frac{4 \pi }{3} & 0  \hdots \\
 0 & 0 & 0 & \frac{4 \pi }{3}  \hdots\\
 \frac{4 \pi }{R_{\text{AB}}} & \frac{2 \sqrt{\frac{2}{3}} e^{-i \alpha _{\text{AB}}} \pi  \text{Sin}\left[\beta _{\text{AB}}\right]}{R_{\text{AB}}^2} & \frac{4 \pi  \text{Cos}\left[\beta _{\text{AB}}\right]}{\sqrt{3} R_{\text{AB}}^2} & -\frac{2 \sqrt{\frac{2}{3}} e^{i \alpha _{\text{\tiny{AB}}}} \thinspace \pi  \text{Sin}\left[\beta _{\text{AB}}\right]}{R_{\text{AB}}^2}  \hdots \\
 -\frac{2 \sqrt{\frac{2}{3}} e^{i \alpha _{\text{AB}}}  \thinspace \pi  \text{Sin}\left[\beta _{\text{AB}}\right]}{R_{\text{AB}}^2} & -\frac{4 \pi  \text{Sin}\left[\beta _{\text{AB}}\right]{}^2}{3 R_{\text{AB}}^3} & -\frac{2 \sqrt{2} e^{i \alpha _{\text{AB}}}  \thinspace \pi \text{Sin}\left[2 \beta _{\text{AB}}\right]}{3 R_{\text{AB}}^3} & \frac{4 e^{2 i \alpha _{\text{AB}}}  \thinspace \pi  \text{Sin}\left[\beta _{\text{AB}}\right]{}^2}{3 R_{\text{AB}}^3}  \hdots \\
 -\frac{4 \pi  \text{Cos}\left[\beta _{\text{AB}}\right]}{\sqrt{3} R_{\text{AB}}^2} & -\frac{2 \sqrt{2} e^{-i \alpha _{\text{AB}}}  \thinspace \pi  \text{Sin}\left[2 \beta _{\text{AB}}\right]}{3 R_{\text{AB}}^3} & -\frac{8 \pi \text{Cos}\left[\beta _{\text{AB}}\right]{}^2}{3 R_{\text{AB}}^3} & \frac{2 \sqrt{2} e^{i \alpha _{\text{AB}}}  \thinspace \pi  \text{Sin}\left[2 \beta _{\text{AB}}\right]}{3 R_{\text{AB}}^3}  \hdots \\
 \frac{2 \sqrt{\frac{2}{3}} e^{-i \alpha _{\text{AB}}}  \thinspace \pi  \text{Sin}\left[\beta _{\text{AB}}\right]}{R_{\text{AB}}^2} & \frac{4 e^{-2 i \alpha _{\text{AB}}}  \thinspace \pi \text{Sin}\left[\beta _{\text{AB}}\right]{}^2}{3 R_{\text{AB}}^3} & \frac{2 \sqrt{2} e^{-i \alpha _{\text{AB}}}  \thinspace \pi   \text{Sin}\left[2 \beta _{\text{AB}}\right]}{3 R_{\text{AB}}^3} & -\frac{4 \pi \text{Sin}\left[\beta _{\text{AB}}\right]{}^2}{3 R_{\text{AB}}^3} \hdots \\
 \frac{4 \pi }{R_{\text{AC}}} & \frac{2 \sqrt{\frac{2}{3}} e^{-i \alpha _{\text{AC}}}  \thinspace \pi  \text{Sin}\left[\beta _{\text{AC}}\right]}{R_{\text{AC}}^2} & \frac{4 \pi  \text{Cos}\left[\beta _{\text{AC}}\right]}{\sqrt{3} R_{\text{AC}}^2} & -\frac{2 \sqrt{\frac{2}{3}} e^{i \alpha _{\text{AC}}}  \thinspace \pi  \text{Sin}\left[\beta _{\text{AC}}\right]}{R_{\text{AC}}^2}  \hdots \\
 \vdots & \vdots & \vdots & \vdots 
\end{array}\right)
$$
As an example, the interaction between the $m=0$ dipole on sphere A and the $m=0$ dipole on sphere B is given by $ \mathbb{K}_{7,3} = -\frac{8 \pi  \text{Cos}\left[\beta _{\text{AB}}\right]{}^2}{3 R_{\text{AB}}^3}$ \thinspace since the seventh term in the vector $\textbf{M}$ is $B_{1}^{0}$ and the third term in $\textbf{M}$ is $A_{1}^{0}$. This term will also appear as $\mathbb{K}_{3,7}$ due to symmetry of the interaction. \\

\bibliographystyle{phjcp}


\begin{thebibliography}{10}

\bibitem{Soules}
{\sc J.~A. {Soules}},
\newblock {\em Am. J. Phys.} {\bf 58}, 1195 (1990).

\bibitem{Slisko}
{\sc J.~{Slisko}} and {\sc R.~A. {Brito-Orta}},
\newblock {\em Am. J. Phys} {\bf 66}, 352 (1998).

\bibitem{Phillips}
{\sc R.~J. {Phillips}},
\newblock {\em Journal of Colloid and Interface Science} {\bf 175}, 386 (1995).

\bibitem{Kwon1998}
{\sc G.~{Kwon}}, {\sc Y.~{Won}}, and {\sc B.~{Yoon}},
\newblock {\em Journal of Colloid and Interface Science} {\bf 205}, 423 (1998).

\bibitem{Shevchenko}
{\sc E.~V. {Shevchenko}},
\newblock {\em Nature} {\bf 439}, 55 (2006).

\bibitem{Talapin}
{\sc D.~V. {Talapin}},
\newblock {\em ACS Nano} {\bf 2}, 109 (2008).

\bibitem{Jackson}
{\sc J.~D. {Jackson}},
\newblock {\em Classical Electrodynamics},
\newblock Wiley, 3rd edition, 1998.

\bibitem{Abramowitz}
{\sc M.~{Abramowitz}} and {\sc I.~A. {Stegun}},
\newblock {\em Handbook of Mathematical Functions with Formulas, Graphs, and
  Mathematical Tables}, chapter~22,
\newblock Dover, 1964.

\bibitem{Messiah}
{\sc A.~{Messiah}},
\newblock {\em Quantum Mechanics}, chapter Vector Addition Coefficients and
  Rotation Matrices,
\newblock Courier Dover Publications, 1999.

\bibitem{Landau}
{\sc L.~D. {Landau}} and {\sc L.~M. {Lifshitz}},
\newblock {\em Quantum Mechanics Non-Relativistic Theory}, chapter~8,
\newblock Butterworth-Heinemann, 1981.

\bibitem{Maxwell}
{\sc J.~C. {Maxwell}},
\newblock {\em A Treatise on Electricity and Magnetism}, volume~1, chapter~11,
\newblock Oxford University Press, 1892.

\end{thebibliography}


\begin{thebibliography}{10}
\bibitem [{Note1()}]{Note1}%
  \BibitemOpen
  \bibinfo {note} {When deriving the form for $E^{BA}$, all odd order
  multipoles need to be multiplied by a $-1$ since the alignment of a positive
  multipole This adds a $(-1)^{\ell +\ell ^{'}}$ factor to the odd ordered B
  multipoles in the derivation of $E^{BA}$.}\BibitemShut {Stop}%
\bibitem [{Note2()}]{Note2}%
  \BibitemOpen
  \bibinfo {note} {This makes the interactions much simpler by eliminating many
  of the azimuthal $m \not =0$ multipoles that occur in the general case
  interaction energy.}\BibitemShut {Stop}%
\bibitem [{Note3()}]{Note3}%
  \BibitemOpen
  \bibinfo {note} {Convention: $E^{AB}_{d-m}$ is the interaction energy between
  a dipole on sphere A and a monopole on sphere B calculated with the $E^{AB}$
  formula. $E^{BA}_{m-q}$ is the interaction energy between a monopole on
  sphere B and a quadrupole on sphere A calculated from the $E^{BA}$
  formula}\BibitemShut {NoStop}%

\bibitem{Soules}
{\sc J.~A. {Soules}},
\newblock {\em Am. J. Phys.} {\bf 58}, 1195 (1990).

\bibitem{Slisko}
{\sc J.~{Slisko}} and {\sc R.~A. {Brito-Orta}},
\newblock {\em Am. J. Phys} {\bf 66}, 352 (1998).

\bibitem{Phillips}
{\sc R.~J. {Phillips}},
\newblock {\em Journal of Colloid and Interface Science} {\bf 175}, 386 (1995).

\bibitem{Kwon1998}
{\sc G.~{Kwon}}, {\sc Y.~{Won}}, and {\sc B.~{Yoon}},
\newblock {\em Journal of Colloid and Interface Science} {\bf 205}, 423 (1998).

\bibitem{Shevchenko}
{\sc E.~V. {Shevchenko}},
\newblock {\em Nature} {\bf 439}, 55 (2006).

\bibitem{Talapin}
{\sc D.~V. {Talapin}},
\newblock {\em ACS Nano} {\bf 2}, 109 (2008).

\bibitem{Jackson}
{\sc J.~D. {Jackson}},
\newblock {\em Classical Electrodynamics},
\newblock Wiley, 3rd edition, 1998.

\bibitem{Abramowitz}
{\sc M.~{Abramowitz}} and {\sc I.~A. {Stegun}},
\newblock {\em Handbook of Mathematical Functions with Formulas, Graphs, and
  Mathematical Tables}, chapter~22,
\newblock Dover, 1964.

\bibitem{Messiah}
{\sc A.~{Messiah}},
\newblock {\em Quantum Mechanics}, chapter Vector Addition Coefficients and
  Rotation Matrices,
\newblock Courier Dover Publications, 1999.

\bibitem{Landau}
{\sc L.~D. {Landau}} and {\sc L.~M. {Lifshitz}},
\newblock {\em Quantum Mechanics Non-Relativistic Theory}, chapter~8,
\newblock Butterworth-Heinemann, 1981.

\bibitem{Maxwell}
{\sc J.~C. {Maxwell}},
\newblock {\em A Treatise on Electricity and Magnetism}, volume~1, chapter~11,
\newblock Oxford University Press, 1892.

\end{thebibliography}

\end{document}